\documentclass[]{aa}  

\usepackage{graphicx}
\usepackage{txfonts}
\usepackage{subcaption}         % necessary for continued figures
\usepackage{lscape}             % to rotate a single page table, example in appendix.
\usepackage{placeins}           % useful with \FloatBarrier, to keep onecolumn floats from drifting to the next section
\usepackage{mathrsfs}                                
\usepackage{enumitem}

\newcommand{\lA}{\lambda_{a}}
\newcommand{\lB}{\lambda_{b}}
\newcommand{\lS}{\lambda_{\text{Sci}}}
\newcommand{\lW}{\lambda_{\text{WFS}}}
\newcommand{\am}{\mathrm{am(\zeta)}}
\newcommand{\Cn}{C_n^2(h)}
\newcommand{\VK}{von K{\'a}rm{\'a}n}

\begin{document}

\title{Chromatic Anisoplanatism in Adaptive Optics}
\subtitle{A detailed comparison of theory with numerical simulations}

\author{B. Femen\'{\i}a-Castell\'a\inst{1}\corrauth{bruno.castella@ohb.de} \and 
        N. Devaney\inst{2}\email{nicholas.devaney@nuigalway.ie} }

\institute{OHB Digital Connect GmbH, Weberstrasse 21, 55130 Mainz, Germany \and 
           Applied Optics Group, Physics Unit, School of Natural Sciences, University of Galway, Ireland}

\date{Received September 1, 2026 / Accepted —}

\abstract
  % context heading (optional)
  %---------------------------
{Chromatic Anisoplanatism arises in Adaptive Optics because atmospheric refraction makes rays of different wavelengths follow different paths through the atmosphere resulting in different wavefronts at the sensing wavelength and the science wavelength, introducing a correction error that grows rapidly with zenith distance.}  
  % aims heading (mandatory)
  %-------------------------
   {Using numerical simulations based purely on geometric optical propagation we aim to validate the theoretical framework describing wavefront errors  caused by Chromatic Anisoplanatism. We characterize the resulting phase variance and Strehl ratio  loss for telescope apertures up to 39~m, retrieving their second-order statistics.} 
  % methods heading (mandatory)
  %----------------------------
   {We compare the analytical expressions for the Chromatic Anisoplanatism wavefront phase variance against independent Monte Carlo simulations. The framework models an 8-layer approximation of the median Cerro Armazones turbulence profile using Kolmogorov and von K\'arm\'an turbulence models. We simulate pupil diameters from 4.2 to 39~m with $\lambda_{\rm WFS} = 589.16$~nm for the wavefront sensing  and $\lambda_{\rm Sci} = 1.25\,\mu\text{m}$ as the science wavelength. A phase-screen recombination strategy increases the number of quasi-independent realizations up to $\sim 5\times10^{5}$, optimizing statistical accuracy.}
  % results heading (mandatory)
  %----------------------------
   {Simulated phase variances match theoretical predictions well across all apertures, though theory slightly overestimates variance in the \VK~case. The Strehl Ratio loss exceeds 30\% for zenith angles~$\gtrsim 70^\circ$. The residual wavefront is dominated by high spatial frequencies; removing tip-tilt reduces phase variance by $\lesssim 1\%$ for apertures $\geq 10$~m. The massive dataset reveals highly non-Gaussian probability density functions for the phase variance and Strehl Ratio values, both characterized by non-zero skewness and  kurtosis, and long tails toward rare, high wavefront phase variance and/or low Strehl Ratio error events.}
  % conclusions heading (optional), leave it empty if necessary
  %------------------------------------------------------------
   {}


\maketitle

\nolinenumbers

%%%%%%%%%%%%%%%%%%%%%%%%%%%%%%%%%%%%%%%%%%%%%%%%%%%%%%%%%%%%%%
\section{Introduction}
\label{sec:intro} 

Adaptive optics (AO) has become an indispensable part of practically all medium to large ground-based telescopes operating at visible and near-infrared wavelengths. Adaptive optics can correct for the effects of atmospheric turbulence and provide diffraction-limited images, thereby dramatically increasing spatial resolution and point detection capability. Since astronomical seeing depends on elevation angle, the performance of AO systems is best at high elevations. However, there are situations where it is required to carry out observations at low elevation angles - this may be the case on large telescopes for objects of great interest which do not culminate at high elevations at the observatory site. Solar telescopes (e.g. see \cite{RM11, DKIST_20, QuinteroC:EST, FC22}) frequently observe at low elevation due to the fact that daytime seeing is minimal near sunrise and sunset. In addition, there is rapidly growing interest in deploying Free Space Optical Communications (FSOC) systems between    Low Earth Orbit (LEO) satellites (which often pass at very low elevation angles, $<10^\circ$) and  Geostationary (GEO)  satellites and the ground in urban and suburban environments where the turbulence conditions are considerably worse than at astronomical observatories.

Chromatic effects (\cite{D08}) can severely limit the AO performance attainable at high elevation angles. These arise from the fact  that the refractive index of air depends on wavelength. Chromatic effects are important in AO systems for two reasons: i) the AO correction is normally over a finite wavelength range, in astronomy usually corresponding to the standard  photometric bands (I,J,H,K etc.) ii) due to the faintness of natural stars, AO systems use laser guide stars (LGS) to increase sky coverage, and these typically operate at the wavelength of the Sodium doublet (589.16 nm) or at even shorter wavelengths in the case of LGS employing Rayleigh backscatter. 

The dispersion of air gives rise to the well-known refractive elongation of images near the horizon. However, this may be corrected using an Atmospheric Dispersion Corrector (ADC), and these are usually employed in AO systems, at least on large telescopes. In addition, the wavefront error is not identical at different wavelengths, giving rise to a chromatic path length error. This can have an effect when a finite correction bandwidth is employed, and the effect increases with telescope diameter. However, it is expected that this effect will be negligible even on large telescopes due to being limited by the outer scale of turbulence \citep{O04}. If an LGS is employed then the effect can be corrected by applying an offset based on a model of the dispersion of air.

This article is concerned with a third effect which we refer to as Chromatic Anisoplanatism (CA). This refers to the displacement due to the dispersion of air of rays of different wavelengths when passing through the atmosphere. In AO, where wavefront measurements are typically carried out at wavelengths shorter than the wavelengths being corrected, this will cause the measured wavefront to be different from the wavefront to be corrected, therefore introducing an error in the correction. The effect is small for the zenith angles at which most observations are carried out ($\zeta \leq 60^\circ$) but it is a rapidly increasing function of zenith angle and may therefore be the limiting factor in some cases.  Different authors \citep{W76, W77, W84, S92, N06, D08} have  studied this error and its implications in astronomical observations but given the low elevations at which CA becomes relevant we think it is plausible that this source of error has been  neglected. Recently \citet{hyde2026anisoplanatic} have  derived closed form expressions for phase variance in dual-wavelength AO, but they consider a constant Kolmogorov turbulence profile, corresponding to horizontal propagation. 

In previous works in \citet{femenia2022} and \citet{d24} we carried out a detailed comparison of the theoretical predictions of \citet{W84, S92} and \cite{N06}. However, as far as we are aware, no exhaustive instrument-independent comparison has been made between theoretical models and numerical simulations. \citet{Motte2024} presented the results of a single case of numerical simulation, corresponding to the RISTRETTO instrument. They refer to chromatic anisoplanatism as ‘Chromatic Pupil Shift’ and calculate it using a plane-parallel atmosphere model. 

In this paper we will carry out the comparison between what we consider the most complete theoretical framework and numerical simulations assuming only geometrical optical propagation across atmospheric turbulence\footnote{Notice this assumption is also made in the  theoretical analysis for the error variance caused by Chromatic Anisoplanatism (CA).}. This makes our simulations  an excellent cross-check for the theoretical expressions of the expected CA phase variance. Beyond this validation of the theory, our numerical simulations are a powerful mechanism to obtain the first and second-order statistics for the phase variance  and the Strehl Ratio (SR) loss due to CA without the need to use any approximation (e.g. the Mar\'echal approximation).

%%%%%%%%%%%%%%%%%%%%%%%%%%%%%%%%%%%%%%%%%%%%%%%%%%%%%%%%%%%%%%
\section{Theoretical framework}
\label{sec:theory} 

Let us denote  $\lS$  the wavelength at which  science is conducted and $\lW$ the wavefront sensing wavelength. Consider two rays at wavelengths $\lS$ and $\lW$  which enter the top of the atmosphere at the same point. Because the atmospheric refraction causes the actual propagation path to be dependent on the wavelength, the rays $\lS$ at $\lW$  land at  different positions on the telescope aperture with a lateral separation denoted by $\Delta b_0(\lA, \lB)$. In the literature, this effect has received multiple names such as  chromatic lateral shift or chromatic displacement.

Now consider the situation faced in AO where we correct the wavefront at $\lS$ on a given location $\vec{\rho}$ on the telescope aperture  with the wavefront at $\lW$ measured at the same $\vec{\rho}$. When tracing back the propagation of  rays at $\lW$ and $\lS$ from $\vec{\rho}$ on the telescope aperture to its source (LGS, NGS, Sun...) we will inevitably discover that at any plane perpendicular to the line-of-sight those  ($\lW, \, \lS)$ rays are laterally displaced (see  Eq.~\eqref{eq:sg2_OPD_CA_2} in the Appendix).

In the context of AO the above difference of ray propagation paths for ($\lW,\,\lS$) implies the accumulated wavefront distributions over the telescope aperture as recorded at $\lS$ and $\lW$ will not be identical and therefore the correction derived from the WFS at  $\lW$ is not   what $\lS$ actually experiences. That is, there is a source of anisoplanatism which we refer to as Chromatic Anisoplanatism (CA). To our knowledge, CA was first analyzed by  \citet{W84} using a $\Delta b_0(\lA, \lB)$ estimator derived for a plane-parallel atmosphere geometry \citep{W76}. Later  works by \citet{S92, N06} independently derived expressions to model the CA but in \citet{femenia2022} we justified why \citet{W84} remains the more rigorous analysis. The theoretical expressions providing the wavefront phase variance caused by CA are summarised in the Appendix together with the  assumptions made and the limitations of this theoretical framework.

More recently  \citet{L22} provided the set of  differential equations rigorously describing the spherical geometry and a numerical method to solve them as well as several approximations. The first order approximation is equivalent to the plane-parallel geometry in \citet{W76}. We have incorporated the spherical geometry in \citet{L22}  in both our theoretical analysis and in the numerical simulations. In \citet{d24} we reported that under typical daytime and nighttime conditions in astronomical observatories, with a weak to medium atmospheric turbulence, when computing the CA effects the approximation of a plane-parallel geometry is excellent for zenith distances $\zeta \leq 60^{\circ}$ and at larger zenith distances  the validity of the approximation starts degrading depending on the  ($\lS$, $\lW$) combination and the atmospheric turbulence strength and vertical distribution.

%%%%%%%%%%%%%%%%%%%%%%%%%%%%%%%%%%%%%%%%%%%%%%%%%%%%%%%%%%%%%%
\section{Numerical Simulations.}
\label{sec:simulations} 

In this section we describe the methodology to perform the  Monte Carlo simulations as well as a series of tactics to reduce numerical error in the results and accelerate the simulations.

%----------------------------------------------
\subsection{The ELT Cerro Armazones turbulence model}
\label{subsec:turbulence_model}

In order to standardise the atmospheric models to be used for simulations of the ELT, \citet{ESO1} provided turbulence models characterised by 35 layers at heights above the telescope from 30~m to 26500~m. The turbulence profiles assume a \VK~turbulence spectrum characterised by an outer scale $L_0=25$~m and while the heights of the turbulence layers are always the same, the different profiles have different integrated and vertical distributions of turbulence. For the purpose of this study we have considered the  median turbulence profile.

\begin{figure}[ht!]
   \centering
   \includegraphics[width=0.95\hsize]{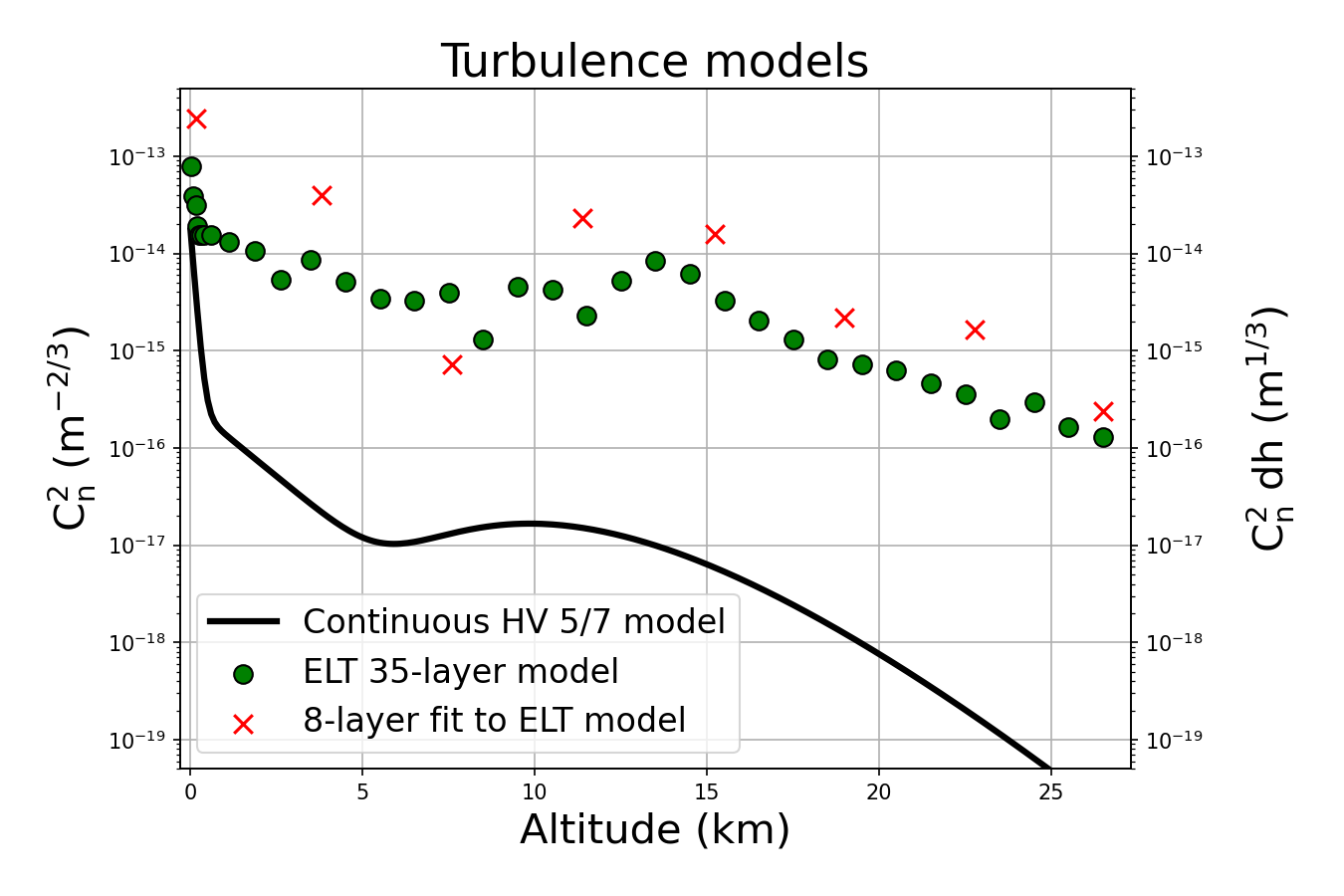}
   \caption{The red crosses show the discrete 8-layer model used in the simulations and derived from the median turbulence 35-layer model for the ELT. As a comparison in solid black line the HV-3/7 model.}
   \label{fig:turb_profile}
\end{figure}

As justified in Sec.~\ref{subsec:PS_generation}, given the extremely high spatial resolution at which the phase screens need to be sampled and the large telescope apertures considered, it would be impractical to simulate a 35-layer model on a consumer laptop (16 GB RAM and no GPU; to our knowledge IDL does not support native GPU compilation out of the box). For this reason we have developed an 8-layer approximation to the 35-layer profile using the approach described in \citet{ellerbroek1994} which performs a fit to the height, and turbulence fraction while preserving the first 16 moments of the original turbulence distributions.

\begin{table}[ht!]
\caption{8-layer turbulence profile in the simulations. }
\label{table:turb_profile}
\centering 
\begin{tabular}{l c c c c} 
\hline\hline
    &           & C$^2_{n,i}$ & Turbulence     & CA fraction \\
$i$ & $h_i$ (m) & (m$^{1/3}$) & fraction $f_i$ &  $w_i$      \\  \hline                      
1   &      172  &   2.441e-13 &     0.7450     &    0.011    \\
2   &     3799  &   3.940e-14 &     0.1203     &    0.208    \\
3   &     7595  &   7.309e-16 &     0.0022     &    0.009    \\
4   &    11377  &   2.338e-14 &     0.0714     &    0.381    \\ 
5   &    15228  &   1.595e-14 &     0.0487     &    0.305    \\
6   &    18987  &   2.177e-15 &     0.0066     &    0.045    \\
7   &    22781  &   1.664e-15 &     0.0051     &    0.036    \\
8   &    26500  &   2.379e-16 &     0.0007     &    0.005    \\  \hline
\end{tabular}
\end{table}

%----------------------------------------------
\subsection{Phase screens generation}
\label{subsec:PS_generation}

Using CAOS \citep{CAOS_2005} we have generated two sets of Phase Sreens (PSs) with Kolmogorov and \VK~turbulence with the characteristics given in Table~\ref{table:PS_characteristics}.  The underlying method is fully described in \citet{carbillet_numerical_2010}. Here we mention that this is an FFT-based method which is well known to underrepresent turbulence associated with sub-pixel spatial scales in the Kolmogorov case (but not for \VK~because of the outer scale). For the Kolmogorov case a set of 6 sub-harmonic additional components are created and added during the generation of each PS to compensate for the turbulence power missed at sub-pixel scales. 

\begin{table}[ht!]
\caption{Characteristics of generated Phase Screens (PSs)}
\label{table:PS_characteristics}
\centering 
\begin{tabular}{l c c } 
\hline\hline
                                     &  Kolmogorov  & \VK          \\  \hline                      
PS Length                            &     41.75 m  &   41.75 m    \\
Nb. Pixels                           &     8192     &   8192       \\
\# Sub-harmonics                     &     6        &   N.A.       \\ 
Pixel sampling                       & $\sim5.1$ mm & $\sim5.1$ mm \\ 
\# pixels for $r_0(\zeta= 0^\circ$ ) &    31        &      31      \\
\# pixels for $r_0(\zeta=75^\circ$ ) &    14        &      14      \\
\# Pupils EPD=4.2 m                  &    81        &      81      \\ 
\# Pupils EPD=10 m                   &    16        &      16      \\
\# Pupils EPD=24 m                   &     1        &       1      \\
\# Pupils EPD=39 m                   &     1        &       1      \\  \hline
\end{tabular}
\tablefoot{The \# Pupils give the number of non-overlapping pupils for each Entrance Pupil Diameter (EPD) that can be fit into a single PS. This is used to increase the number of simulation realizations.}
\end{table}

The  PS length of 41.75 m is given by the need to accommodate the ELT aperture diameter plus some margin to shift the pupil footprint at each turbulent layer height to represent the differential lateral displacement between $\lW$ and $\lS$. The pixel size is needed to at least capture with one pixel such lateral differential shifts at the lowest turbulent layer in  those $\zeta$ cases where CA is expected to be noticeable. 

The PSs are the most computationally intensive part of the numerical simulation and so it makes sense to compute them only once and store them rather than generating them dynamically whenever a simulation is executed for different conditions (e.g. $\zeta$ in the simulations presented in this work but also to check other ($\lW, \lS$) cases). For both the Kolmogorov and \VK~scenarios we precomputed and saved to disk a set of 6400 statistically independent PSs. The PSs were generated with an intrinsic $r_0(0.5\mu m)=1$~m towards zenith ($\zeta = 0$, airmass is 1); we refer to this generated PS as: $\mathrm{PS}^{(k)}(r_0=1~\mathrm{m}, \zeta=0)$ .

When we generate a realization of 8 turbulent layers  we combine 8 different  $\mathrm{PS}^{(k)}(r_0=1~\mathrm{m}, \zeta=0)$; for realization $m$ in the simulation the atmospheric turbulence  considers the combination of PSs defined by the 8-element tuple $\pi_{m}$ (see Section ~\ref{subsec:tuple_generation}). Each of these eight selected PSs is renormalized in amplitude to represent the actual turbulence in Table~\ref{table:turb_profile} and the actual zenith distance $\zeta$ to be simulated. Explicitly, if in $\pi_{m}$ for layer $i$ with turbulence fraction $f_i$ we use  Phase Screen $\mathrm{PS}^{(k)}$ and  airmass am($\zeta$) towards zenith distance $\zeta$ then the nominal $\mathrm{PS}^{(k)}$ is scaled according to:

\begin{equation}
  \label{eq:PS_normalize}
  \mathrm{PS}_{i}^\prime(f_i, r_0, \zeta) = \sqrt{f_i \cdot \mathrm{am}(\zeta)} \cdot r_0^{5/6} \cdot \mathrm{PS}^{(k)}(r_0=1~\mathrm{m}, \zeta=0)
\end{equation}
   
For the computation of am($\zeta$) we use the approximation in \citet{Kasten1989} which for the ELT site gives an error less than 1\% for $\zeta \leq 87^{\circ}$ with respect to the actual numerical computation of the geometric airmass. 

Care is needed when comparing results for Kolmogorov and \VK~turbulence since renormalization of a Kolmogorov PSD into a \VK~PSD is a mathematically ill-posed problem. We show results where both the Kolmogorov and \VK~phase screens are computed with the same value of $r_0=0.157m$ as well as results where the Kolmogorov PS are computed for $r_0^\mathrm{Kolmo}(0.5\mu m)=0.183$~m which gives the same phase variance as the \VK~case with $r_0(0.5\mu m)=0.157$~m and $L_0=25$~m over a pupil diameter of 0.157~m.

%----------------------------------------------
\subsection{Increasing the number of instances}
\label{subsec:tuple_generation}

\begin{figure*}[ht!]
  \centering
  \resizebox{17cm}{8.5cm}
    {\includegraphics{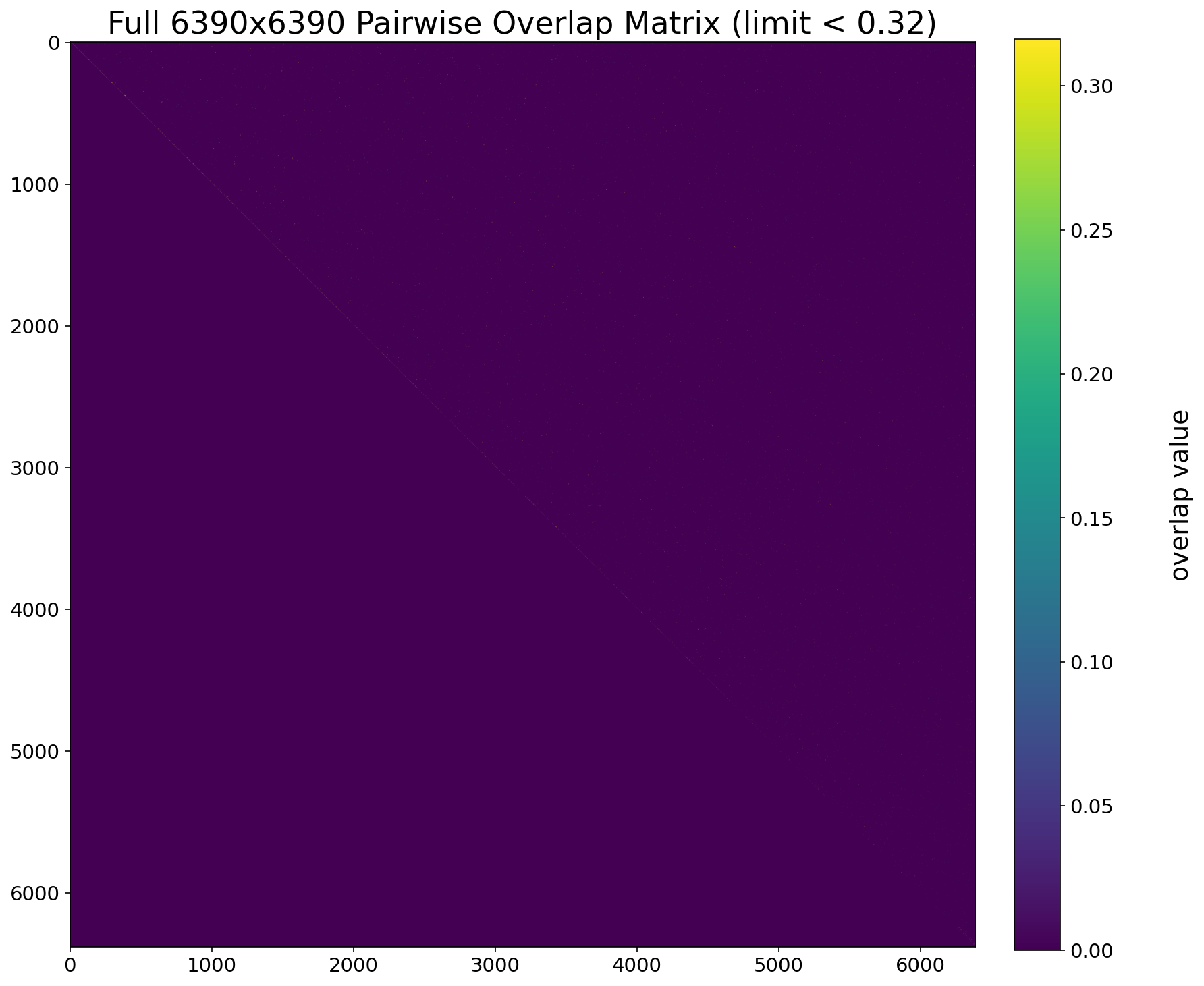}
     \includegraphics{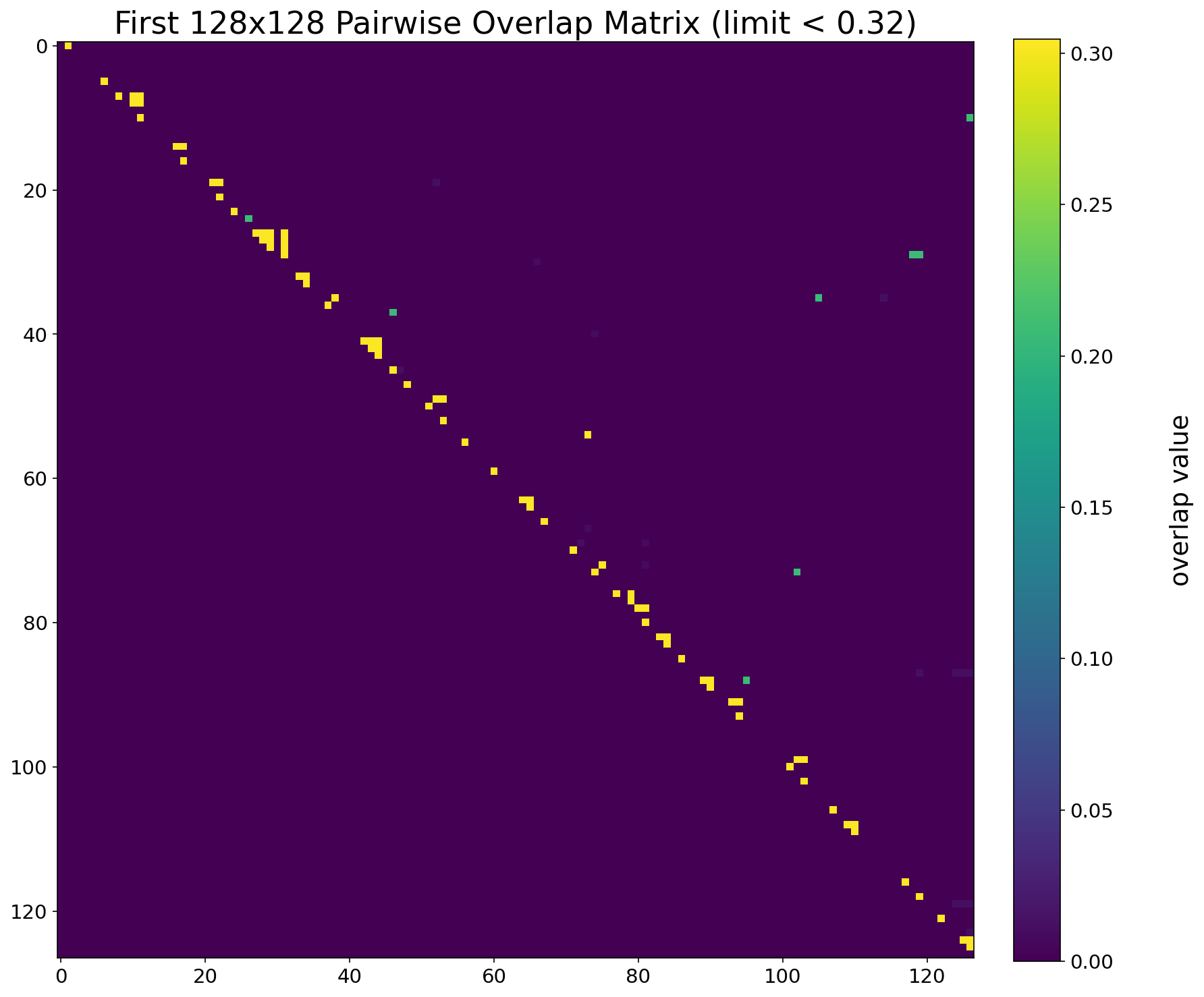}}
     \caption{Pair-wise overlapping metric between any of the 6386 combinations of 8 layers. \textbf{Left:} full matrix view shows it is extremely sparse. \textbf{Right:} zooming the first $128\times128$ quadrant allows to see the rare combinations where the metric is above 1\%.}
  \label{fig:tuples_overlap}
\end{figure*}

At each height $h_i$ there is a different turbulence fraction $f_i$ which results in each layer having a different contribution to the total wavefront phase variance caused by CA as shown  in the last column in Table~\ref{table:turb_profile}. We see that although the first turbulence layer at $h_1=172$~m contributes almost 75\% to the total  turbulence power, this layer is responsible for only 1.1\% of the CA phase variance. This is because the lateral separation between the pupil footprints at $\lW$ and $\lS$ is very small at the lowest layer \footnote{Similar to the angular anisoplanatism,  the effect would be zero if all the turbulence is located at the telescope entrance pupil.}. On the other hand we notice that a small set of layers at high altitudes having minor contributions to the turbulence power are responsible for most of the CA phase variance: layers  $i=2,$ 4 and 5 (at heights at 3799, 11377 and 15228~m above the telescope, respectively) account for $\simeq 24\%$  of the total turbulence power but are responsible for 90\% of the total CA phase variance.  This gives us a method to  significantly increase the number of simulated PS combinations if we are willing to include weakly correlated as well as completely decorrelated atmospheric turbulence instances.

Let us denote with $\pi_m$ the 8-element tuple with the indexes of the set of 8 $\mathrm{PS}^{(k)}$ to build the turbulence instance in iteration $m$ of the simulation. As an example consider the tuples $\pi_m=(1,2,3,4,5,6,7,8)$ and $\pi_n=(9, 10,  11, 1, 12, 13, 14, 15)$ corresponding to two different iterations $m$ and $n$ of the simulation; the turbulence realizations in those iterations are nearly independent of each other except that $\mathrm{PS}^{(1)}$ is at layer $i$=1 in $\pi_m$  and  at layer $i=4$ in $\pi_n$. We then expect that the results from iterations $m$ and $n$ will be residual wavefronts caused by CA that are almost completely decorrelated and at most there should be a very small correlation of around 1.1\% \footnote{The value of 1.1\% is the minimum between $w_1$ and $w_4$ in Table~\ref{table:turb_profile}.}. Notice that a larger decorrelation can be achieved by rotating by multiples of $90^{\circ}$ and/or flipping and/or mirroring and/or sign inverting $\mathrm{PS}^{(1)}$ between $\pi_m$ and $\pi_n$.  We can extend this rationale for $\pi_m$ and $\pi_n$ sharing multiple  $\mathrm{PS}^{(k)}$ by computing an "overlap" metric which is  the CA fraction weight-penalized measure of common elements between two sequences  $\pi_m$ and $\pi_n$ based on their index positions within the tuples:

\begin{equation}
  \text{Overlap}(\pi_m, \pi_n) = \sum_{k \in \pi_n \cap \pi_m} \min\Big( w\big[\text{idx}(\pi_m, k)\big], \, w\big[\text{idx}(\pi_n, k)\big] \Big)
\end{equation}

\noindent where idx($\pi_m,k$) denotes the index position of element $k$  within the tuple $\pi_m$. In short, this metric scores how much information two sequences $\pi_m$ and $\pi_n$ share. An element contributes to the score only if it exists in both sequences and then its contribution is bounded by the minimum weight of the positions occupied across both sequences. 

Restricting to completely independent $\pi_m$ sequences would have resulted to 800 instances in our simulation. By permitting a maximum overlap of up to 32\%, the number of distinct, usable instances increases to 6386. In any case, the maximum overlap of 32\% rarely happens as illustrated in Fig.~\ref{fig:tuples_overlap}: on the left image we show the full $6386 \times 6386$ overlap matrix of the selected sequences and on the right image we zoom into the narrow diagonal band to be able to see any relevant overlap above 1\%.

%----------------------------------------------
\subsection{Building the simulation}
\label{subsec:simulation_shifting}

\begin{figure*}[ht!]
  \centering
  \includegraphics[width=\hsize]{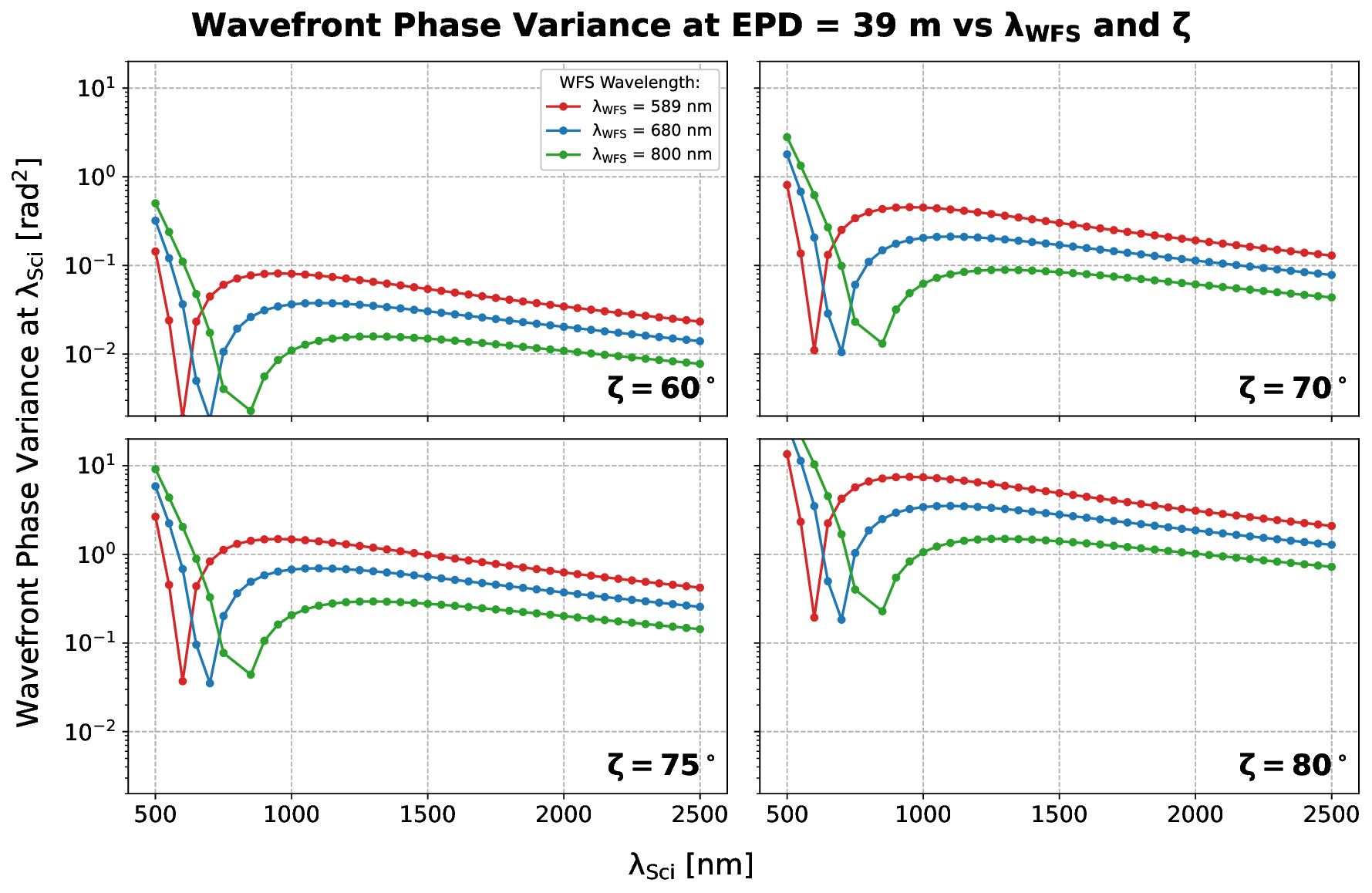}
  \caption{Wavefront phase variance in [rad$^2$] caused by CA for EPD=39~m. The color lines correspond to different values for $\lW$.} 
  \label{fig:CA_vs_lSci}
\end{figure*}

The CA error depends on the particular choice of $\lW$ and $\lS$ as shown in Fig.~\ref{fig:CA_vs_lSci} where we plot for a 39~m telescope aperture the wavefront phase variance values obtained for different $\lW$ in the WFS. This kind of plot is easily obtained from the theoretical framework and could be considered in the early stages of system design if intended to be operated at low elevations.    

However, in this work we are concerned with crosschecking the validity of these theoretical expressions with independent numerical simulations and also  obtaining additional valuable information which cannot be retrieved from the theory. For this crosscheck we  choose $\lW=589.16$~nm and $\lS=1.25~\mu$m corresponding to a common scenario in nighttime AO with a Sodium LGS to increase sky coverage. A specific case of telescope aperture and $(\lW, \lS, \zeta)$ can be easily simulated on demand at the expense of a  large execution time (e.g. in our laptop computing the theoretical expression for an scenario takes 0.5 seconds; the simulation would require about 1 week to run). 

\begin{figure}[ht!]
  \centering
  \includegraphics[width=\hsize]{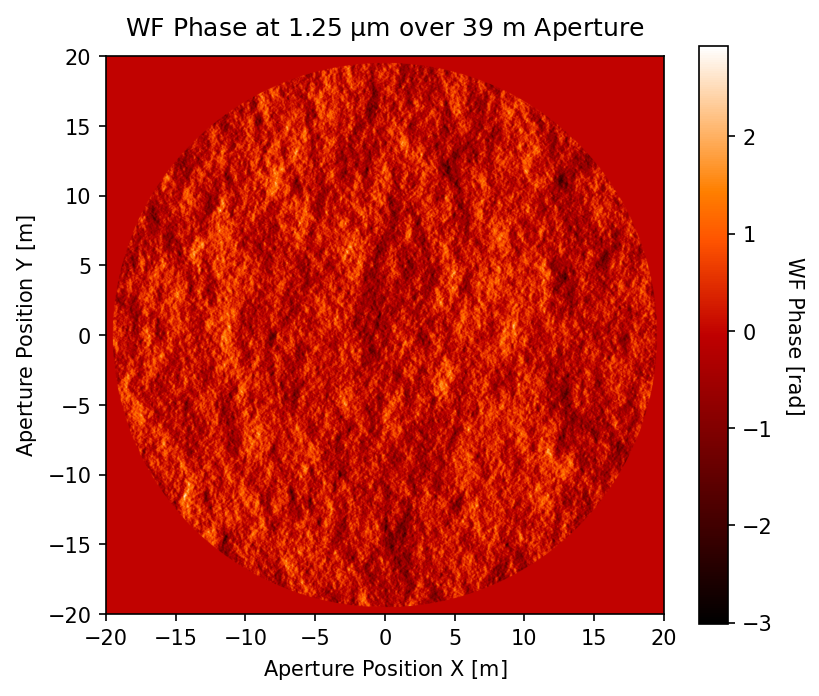}
  \caption{Example of a residual WF phase at 1.25~$\mu$m when the WFS operates at 589.16~nm for an aperture size of 39~m  using \VK~turbulence looking towards $\zeta=70^{\circ}$.} 
  \label{fig:wf_example}
\end{figure}

Our simulations consider telescope apertures 4.2, 10, 24 and 39~m in diameter (to which we also refer to as Entrance Pupil Diameter (EPD). Conceptually the simulation is very easy to understand. Once the PSs are generated using CAOS \citep{CAOS_2005} and arranged to produce an instance of turbulence (see Sect.~\ref{subsec:PS_generation} and \ref{subsec:tuple_generation}) the whole simulation proceeds outside CAOS and can be summarised in a few steps as follows:

\begin{enumerate}
  \item Generate the wavefront as seen on the telescope aperture at $\lW$: $W(\vec{\rho},\lW)$ is the wavefront at position $\vec{\rho}$. For the light propagation through atmospheric turbulence we use the near-field approximation whereby the phase at the telescope aperture is the addition of phase from the different layers intercepted by the rays as they propagate. The near-field approximation is also the mechanism implemented in CAOS, where \citet{CAOS_2005} refer to it as "geometric propagation" . In the Appendix we justify this approach.  
  
\item We do the same to compute $W(\vec{\rho},\lS)$ for which each turbulent layer is shifted due to the effects of atmospheric refraction according to the lateral shift in Eq.~\eqref{eq:sg2_OPD_CA_2}. See later in the main text for the details on how lateral shifts are introduced.

\item The CA error in OPD at the telescope pupil is: $\Delta W(\vec{\rho},\lS, \lW)= W(\vec{\rho},\lS) - W(\vec{\rho},\lW)$. This is what we generate at each of our nearly independent 6386 simulation iterations. Figure~\ref{fig:wf_example} is an example of such a residual wavefront for the case of $\zeta=70^{\circ}$ for an EPD= 39~m showing that the CA-induced residual wavefront is dominated by high spatial frequencies.

\item For each iteration  the wavefront phase at $\lS$ is $\phi(\vec{\rho}) = \Delta W(\vec{\rho},\lS, \lW)\cdot 2 \pi /\lS$ from which we can compute its variance by integrating over each entrance pupil diameter considered (EPD=4.2, 10, 24 and 39~m) and we compute the Strehl Ratio (SR) using its mathematical definition, Eq.~\eqref{eq:SR_def}:
\begin{equation}
  \text{SR}(\lS) = \frac{\left|\displaystyle\int_{\mathcal{A}} e^{i\phi(\vec{\rho})}\,d^2\vec{\rho}\right|^2}
           {\left|\displaystyle\int_{\mathcal{A}} d^2\vec{\rho}\right|^2}
    = \left|\frac{1}{A}\int_{\mathcal{A}} e^{i\phi(\vec{\rho})}\,d^2\vec{\rho}\right|^2
  \label{eq:SR_def}
\end{equation}

\end{enumerate}

An important aspect of the simulation code is to ensure that at a given zenith distance $\zeta$ the lateral shifts induced by the atmospheric refraction correspond to an integer number of pixels. There are two main reasons for this: (i) fractional pixel shifts would require interpolation techniques with the potential to distort the actual CA effects (ii) interpolating a $8192 \times 8192$ array is much more computationally expensive than an array shift. 

To ensure an integer pixel shift at each turbulent layer, the nominal heights at the second column in Table~\ref{table:turb_profile} are modified; an example is shown in Table~\ref{table:new_h} for the case $\zeta=74^{\circ}$, which also shows the lateral separation at each layer for the nominal heights in Table~\ref{table:new_h} and those finally used in the simulation. Notice that the effect of the airmass is taken into account in the computation of the lateral separation \citep{L22} as implemented in Eq.~\eqref{eq:sg2_OPD_CA_2}. The effect of the "perturbed" turbulence layer's height is minimal as shown in Table~\ref{table:new_h_effect} which using the theoretical expressions in the Appendix compute the phase variance for the actual heights in the simulation for each $\zeta$ scenario vs the nominal heights of the model in Table~\ref{table:turb_profile}.

\begin{table}[ht!]
\caption{Nominal vs. perturbed heights to ensure integer pixel shifts for the $\zeta=74^{\circ}$ case. The simulation spatial resolution is $~\simeq 5.1$ mm/pixel.}
\label{table:new_h}
\centering 
\begin{tabular}{c c c c c} 
\hline\hline
 Nominal   & Lateral   & Lateral     & Modified  & Lateral      \\                   
 $h_i$ (m) & Shift (m) & Shift (pix) & $h_i$ (m) & Shift (pix)  \\  \hline                      
     172   &  0.0055   &      1.08   &     160   &     1        \\
    3799   &  0.1006   &     19.74   &    3862   &    20        \\
    7595   &  0.1660   &     32.57   &    7756   &    33        \\
   11377   &  0.2050   &     40.23   &   11224   &    40        \\ 
   15228   &  0.2268   &     44.51   &   15840   &    45        \\
   18987   &  0.2385   &     46.80   &   19432   &    47        \\
   22781   &  0.2452   &     48.11   &   22325   &    48        \\
   26500   &  0.2486   &     48.78   &   28501   &    49        \\ \hline
\end{tabular}
\end{table}

\begin{table}[ht!]
\caption{Relative \% difference when computing the theoretical CA phase variance between the nominal and the perturbed heights for each $\zeta$ scenario.}
\label{table:new_h_effect}
\centering 
\begin{tabular}{l c c c c} 
\hline\hline
 $\zeta$      &  D= 4.2 m  & D= 10 m   & D= 24 m   & D= 39 m  \\  \hline                      
 60$^{\circ}$ &    -2.11   &    -2.15  &   -2.18   & -2.20    \\       
 64$^{\circ}$ &    -3.48   &    -3.41  &   -3.36   & -3.35    \\       
 68$^{\circ}$ &     0.24   &     0.14  &    0.08   &  0.05    \\       
 70$^{\circ}$ &     2.70   &     2.67  &    2.66   &  2.65    \\       
 72$^{\circ}$ &     0.93   &     0.89  &    0.86   &  0.86    \\       
 74$^{\circ}$ &     0.47   &     0.50  &    0.51   &  0.52    \\       
 75$^{\circ}$ &    -0.96   &    -0.93  &   -0.91   & -0.90    \\       
 80$^{\circ}$ &     0.56   &     0.50  &    0.46   &  0.45    \\ \hline
\end{tabular}
\end{table}

An additional optimization is achieved by noticing that for the EPD=4.2 and 10~m we can fit multiple non-overlapping pupils on the same $8192 \times 8192$ pixel PS (i.e. the physical screen size is $41.75$~m, see Table~\ref{table:PS_characteristics}). Because  these fitted  telescope apertures are laterally displaced with respect to each other by a much longer distance than any of the lateral separations in Table~\ref{table:new_h} or the equivalent $r_0$ at each layer height, they can be regarded as uncorrelated instances. This allows to increase the number of simulation results from 6386 iterations to 102176 and 517266 values for EPD=10~m and EPD=4.2~m, respectively.

%%%%%%%%%%%%%%%%%%%%%%%%%%%%%%%%%%%%%%%%%%%%%%%%%%%%%%%%%%%%%%
\section{Simulation results and discussion}
\label{sec:results}

\begin{figure*}[ht!]
    \centering
    \begin{subfigure}{0.9\textwidth}
        \centering
        \includegraphics[width=\textwidth]{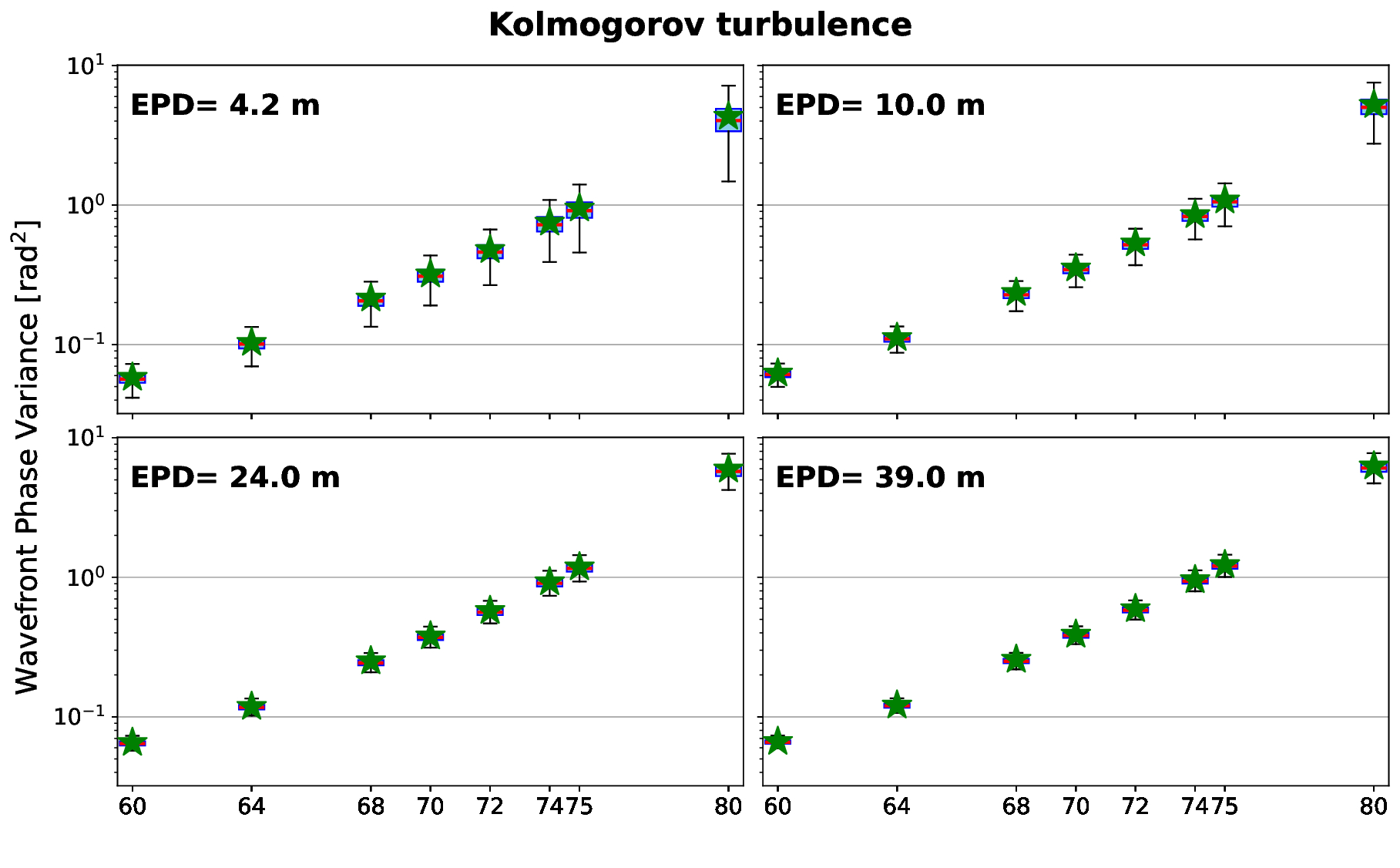}
    \end{subfigure}

    \begin{subfigure}{0.9\textwidth}
        \centering
        \includegraphics[width=\textwidth]{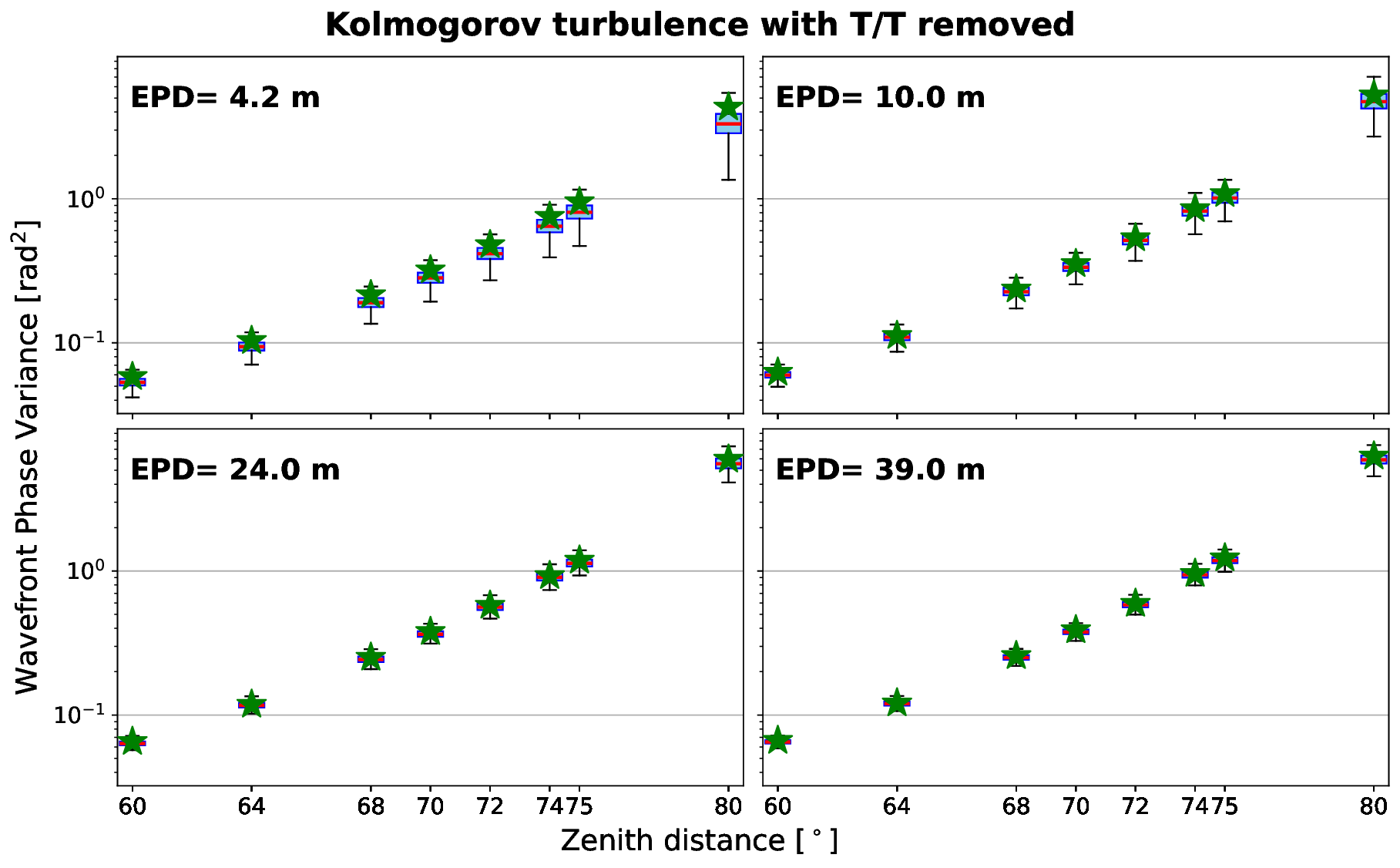}
    \end{subfigure}
    
    \caption{Wavefront phase variance due to CA evaluated at the science wavelength ($\lambda_{\rm Sci} = 1.25\,\mu\text{m}$) as a function of zenith distance ($\zeta$) for entrance pupil diameters (EPD) of 4.2, 10, 24, and 39~m. The wavefront sensor operates at $\lambda_{\rm WFS} = 589.16$~nm. The top panel shows the full \textbf{Kolmogorov turbulence case} and the bottom panel shows the Kolmogorov case with the CA-induced tip-tilt components filtered out. The green star symbols represent the analytical values derived from the complete Kolmogorov theory. The outliers have been omitted to optimize visual clarity.}
    \label{fig:boxplot_sg2_Kolmo}
\end{figure*}

\begin{figure*}[ht!]
    \centering
    \begin{subfigure}{0.9\textwidth}
        \centering
        \includegraphics[width=\textwidth]{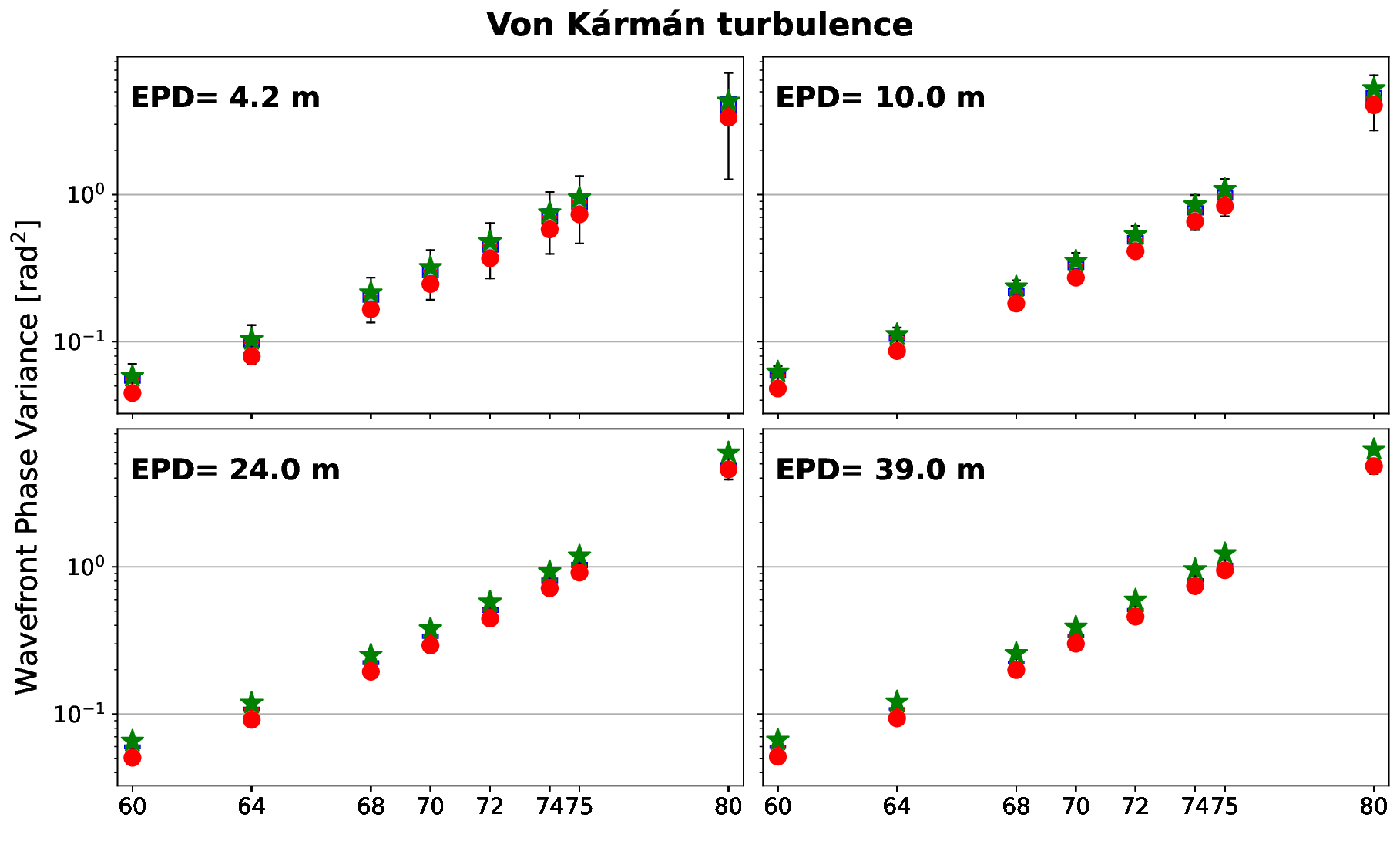}
    \end{subfigure}
    
    \begin{subfigure}{0.9\textwidth}
        \centering
        \includegraphics[width=\textwidth]{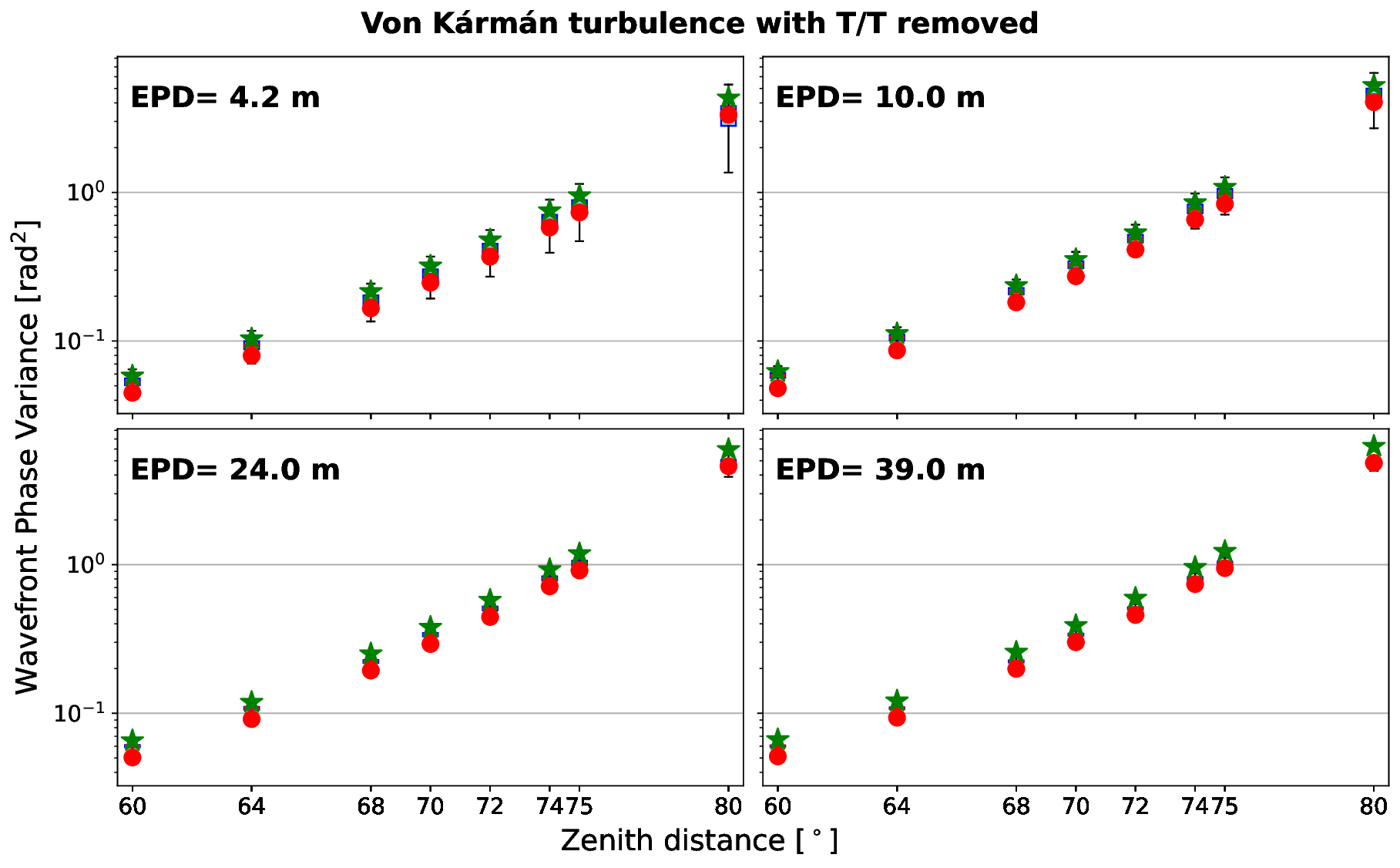}
    \end{subfigure}
    
    \caption{Wavefront phase variance due to CA evaluated at the science wavelength ($\lambda_{\rm Sci} = 1.25\,\mu\text{m}$) as a function of zenith distance ($\zeta$) for entrance pupil diameters (EPD) of 4.2, 10, 24, and 39~m. The wavefront sensor operates at $\lambda_{\rm WFS} = 589.16$~nm. The top panel shows the full \textbf{\VK~turbulence case} and the bottom panel shows the  \VK~case with the CA-induced tip-tilt components filtered out. The green star symbols represent the analytical values derived from the complete Kolmogorov theory, serving as a baseline for cross-comparison. The solid red circles denote the theoretical values adjusted for an equivalent Kolmogorov Fried parameter ($r_0 = 0.183$~m), see  Section~\ref{subsec:PS_generation}. The outliers have been omitted to optimize visual clarity.}
    \label{fig:boxplot_sg2_VK}
\end{figure*}

\begin{figure*}[ht!]
    \centering
    \begin{subfigure}{0.495\textwidth}
        \centering
        \includegraphics[width=\textwidth]{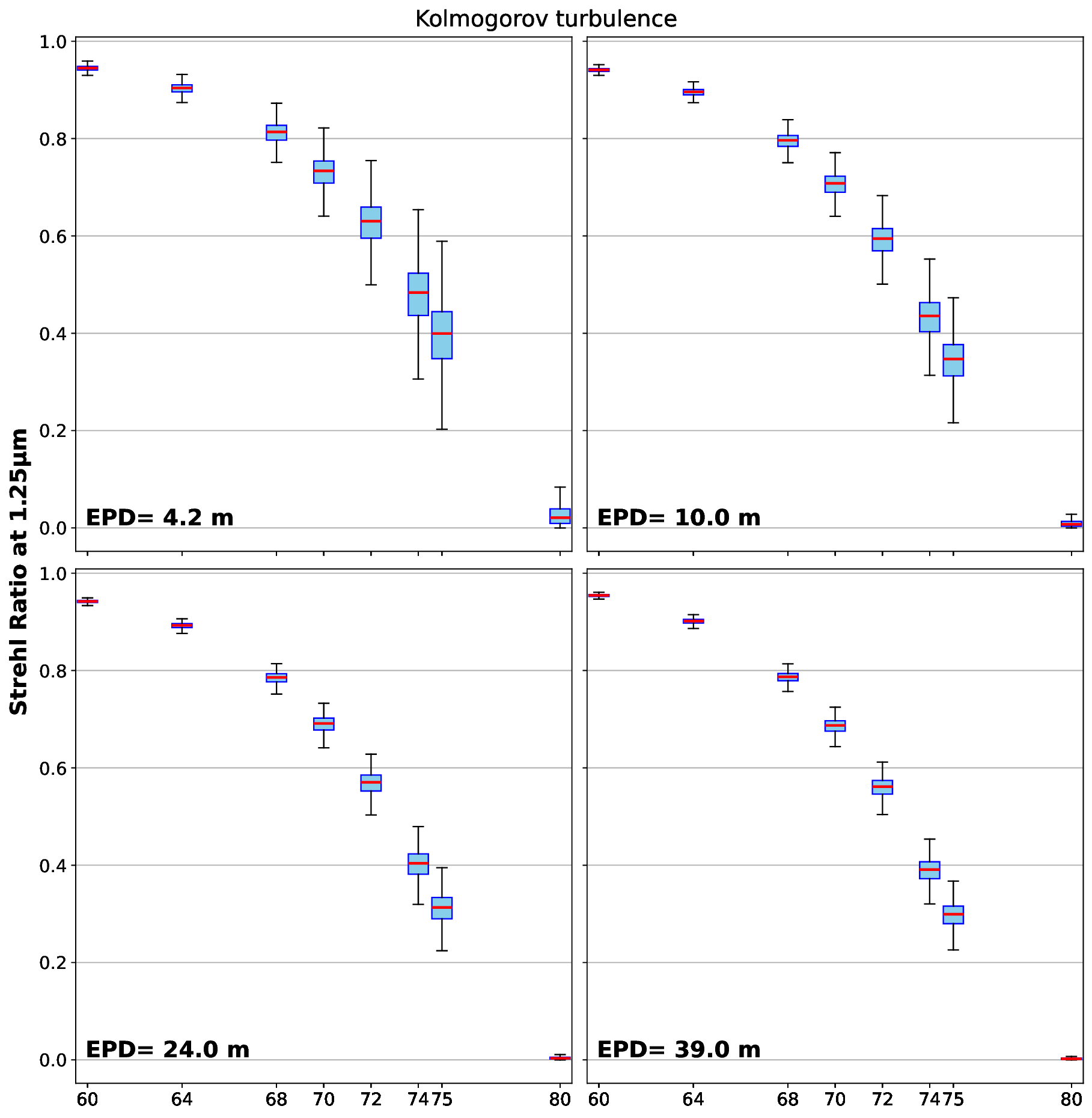}
    \end{subfigure}
    \hfill
    \begin{subfigure}{0.495\textwidth}
        \centering
        \includegraphics[width=\textwidth]{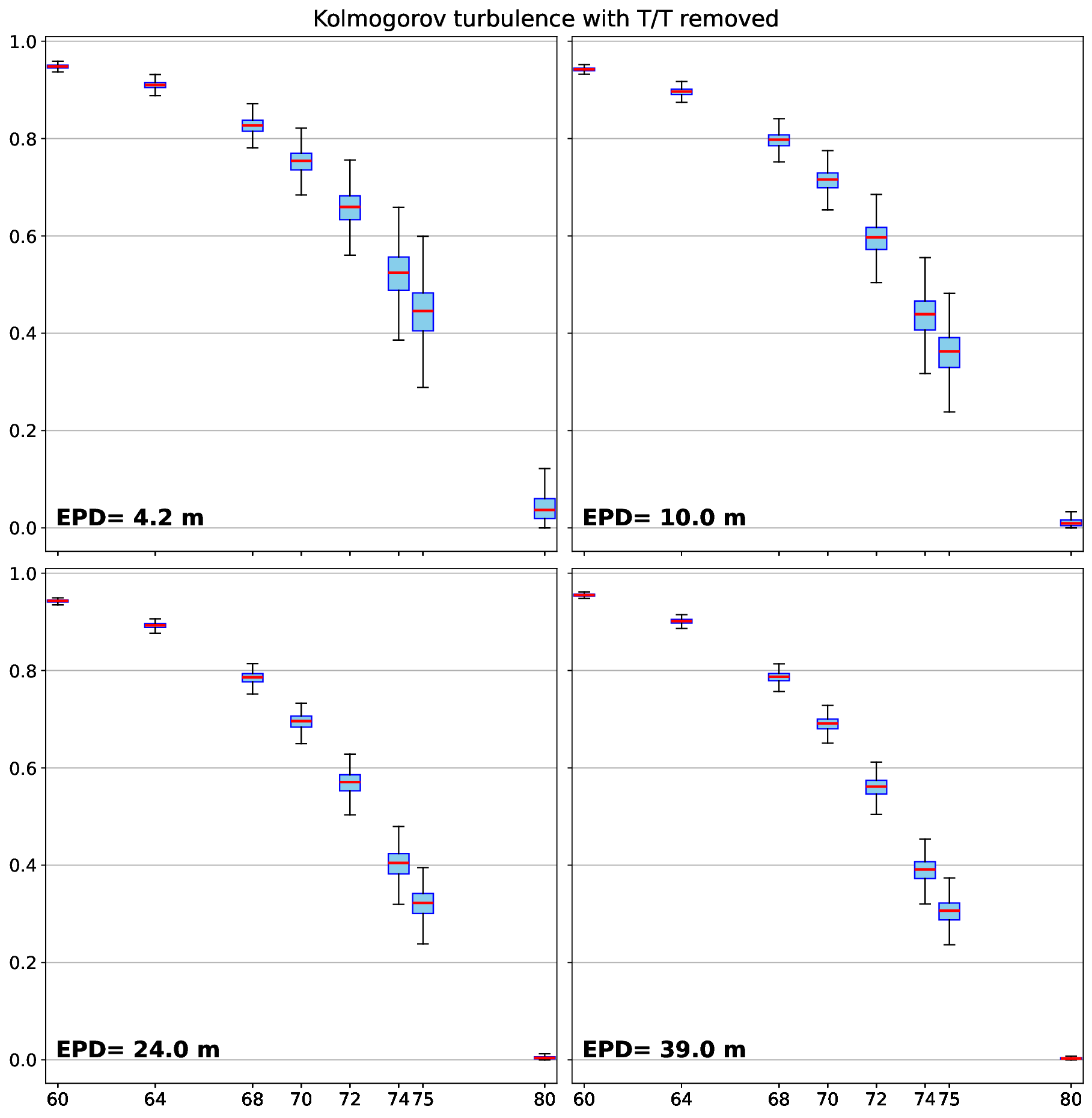}
    \end{subfigure}

    \begin{subfigure}{0.495\textwidth}
        \centering
        \includegraphics[width=\textwidth]{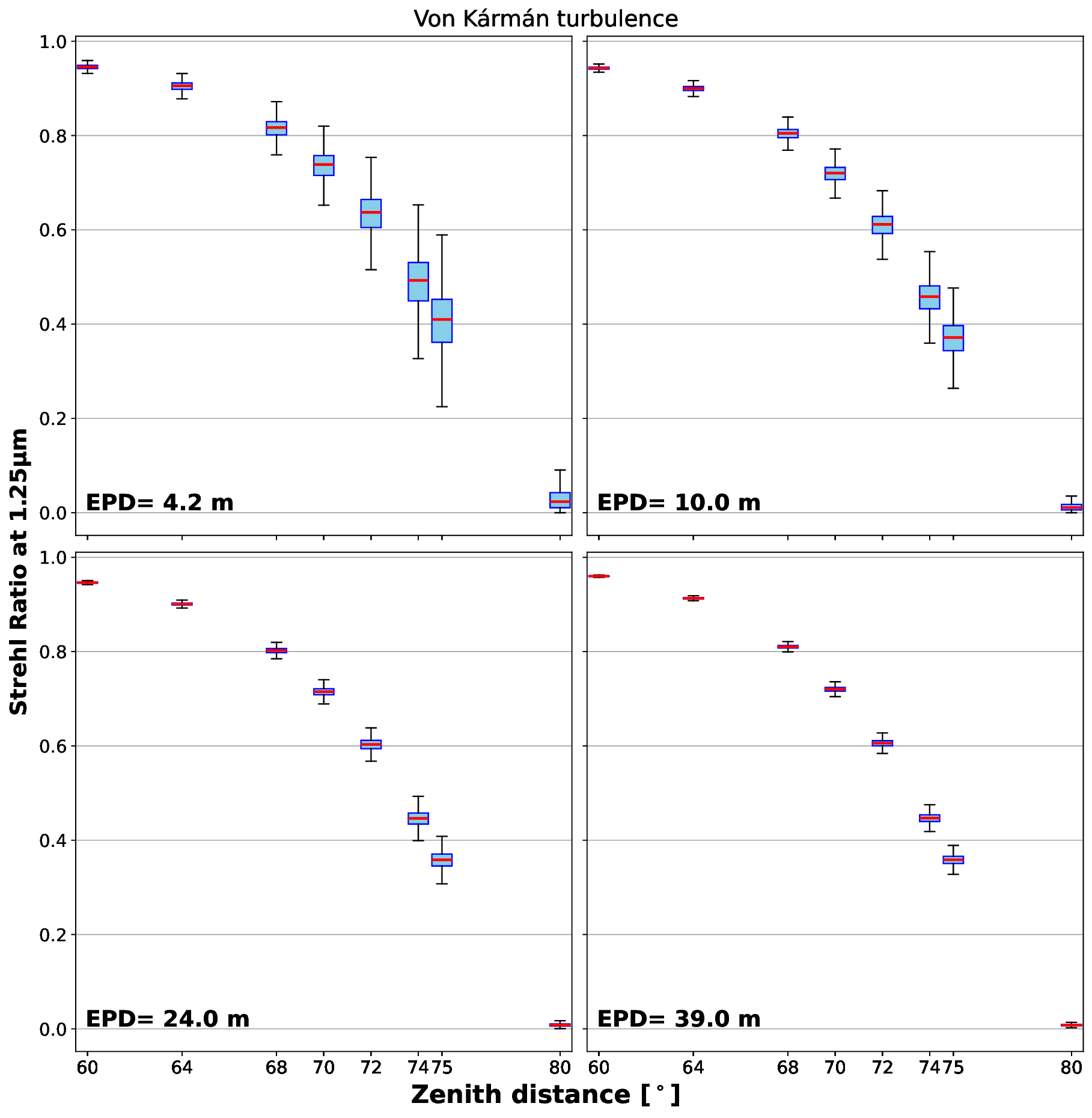}
    \end{subfigure}
    \hfill
    \begin{subfigure}{0.495\textwidth}
        \centering
        \includegraphics[width=\textwidth]{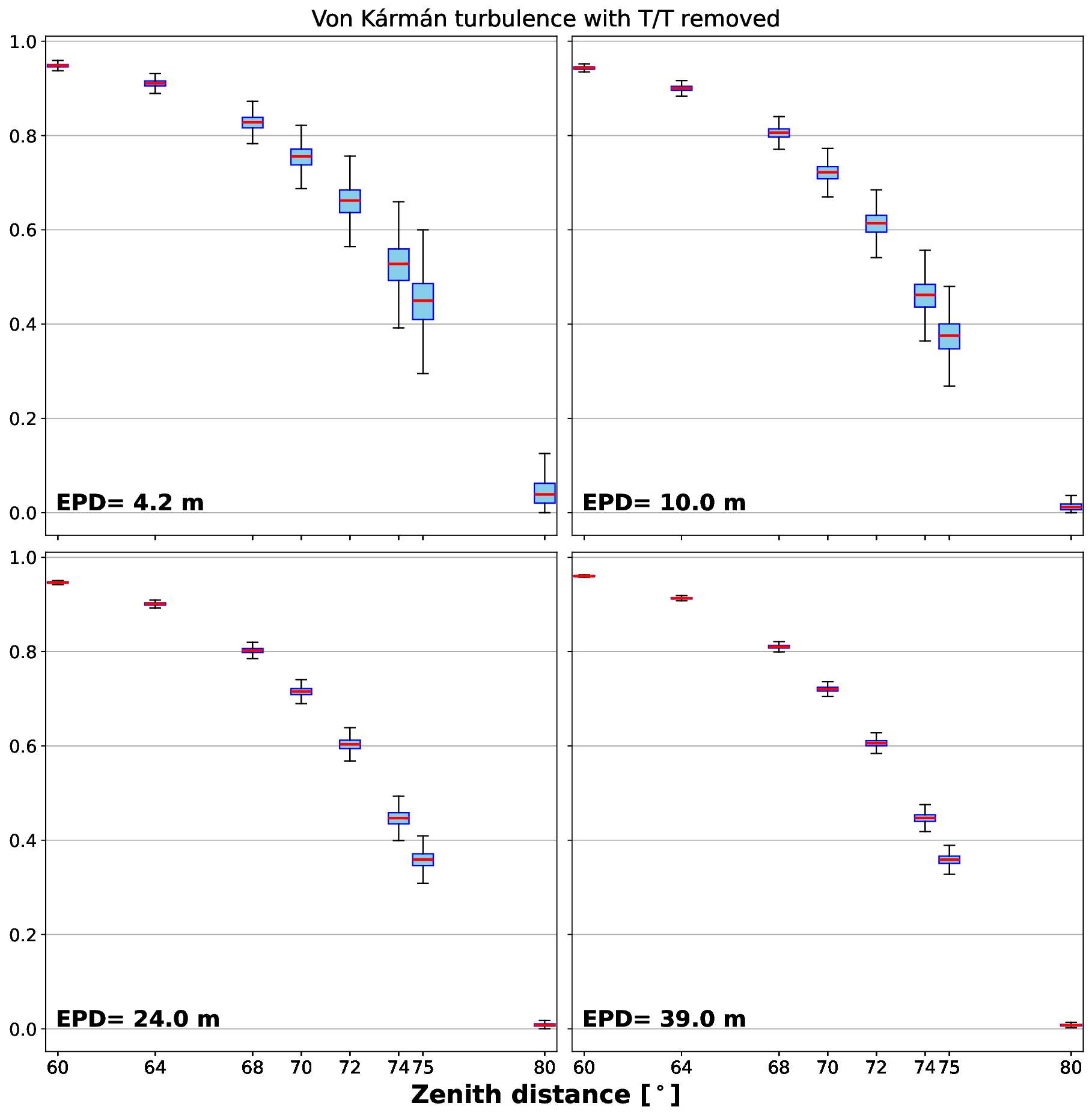}
    \end{subfigure}
\caption{SR values evaluated at the science wavelength ($\lambda_{\rm Sci} = 1.25\,\mu\text{m}$) as a function of zenith distance ($\zeta$) across telescope entrance pupil diameters (EPD) of 4.2, 10, 24, and 39~m, with wavefront sensing conducted at $\lambda_{\rm WFS} = 589.16$~nm. The top panels show the resulting SR for the full Kolmogorov turbulence profile (top-left) and the Kolmogorov profile with the CA-induced tip-tilt component filtered out (top-right). The bottom panels present the corresponding full von K\'arm\'an turbulence scenarios ($L_0 = 25$~m) before (bottom-left) and after (bottom-right) tip-tilt removal. The trends track the rapid degradation of final image quality at large zenith angles ($\zeta \gtrsim 70^\circ$), demonstrating severe performance loss as the aperture scale increases under Kolmogorov conditions. Statistical outliers have been omitted to ease visual comparison across the configurations.}    
    \label{fig:boxplot_SR}
\end{figure*}

\begin{sidewaysfigure*}
    \centering
    \begin{subfigure}{0.495\textwidth}     % Use {0.495\textwidth} for non-REFEREE version
        \centering
        \includegraphics[width=\textwidth]{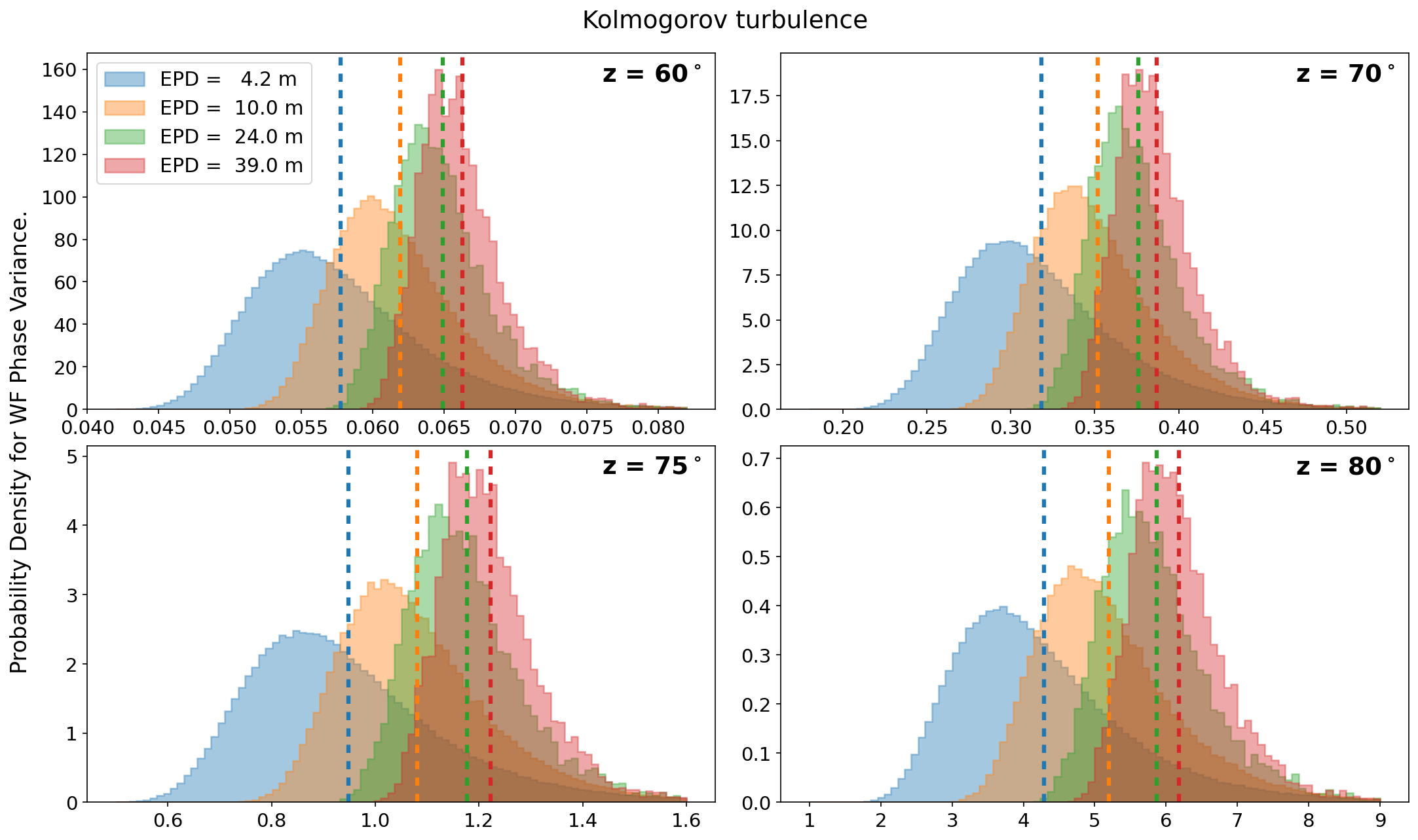}
    \end{subfigure}
    \hfill
    \begin{subfigure}{0.495\textwidth}     % Use {0.495\textwidth} for non-REFEREE version
        \centering
        \includegraphics[width=\textwidth]{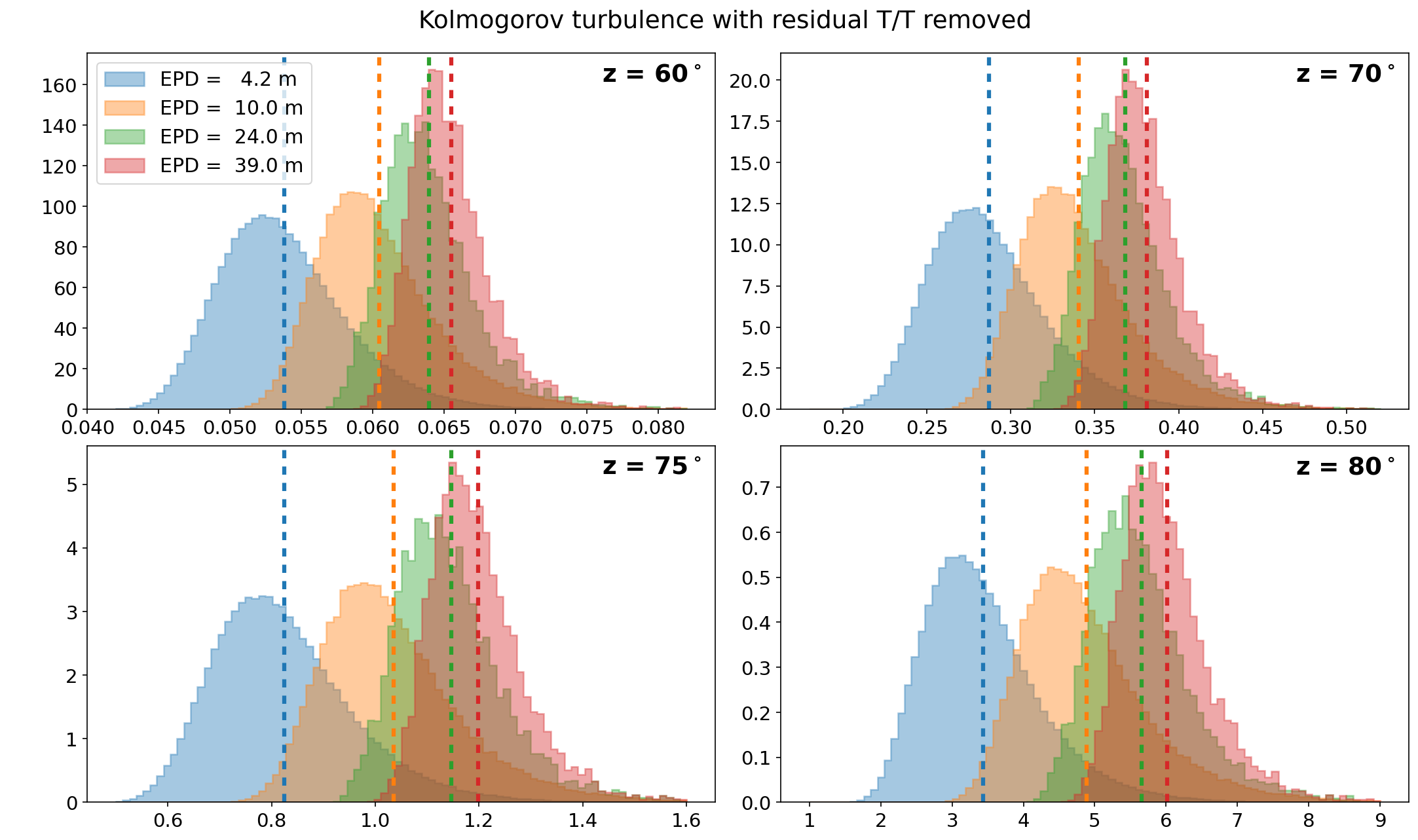}
    \end{subfigure}

    \begin{subfigure}{0.495\textwidth}     % Use {0.495\textwidth} for non-REFEREE version
        \centering
        \includegraphics[width=\textwidth]{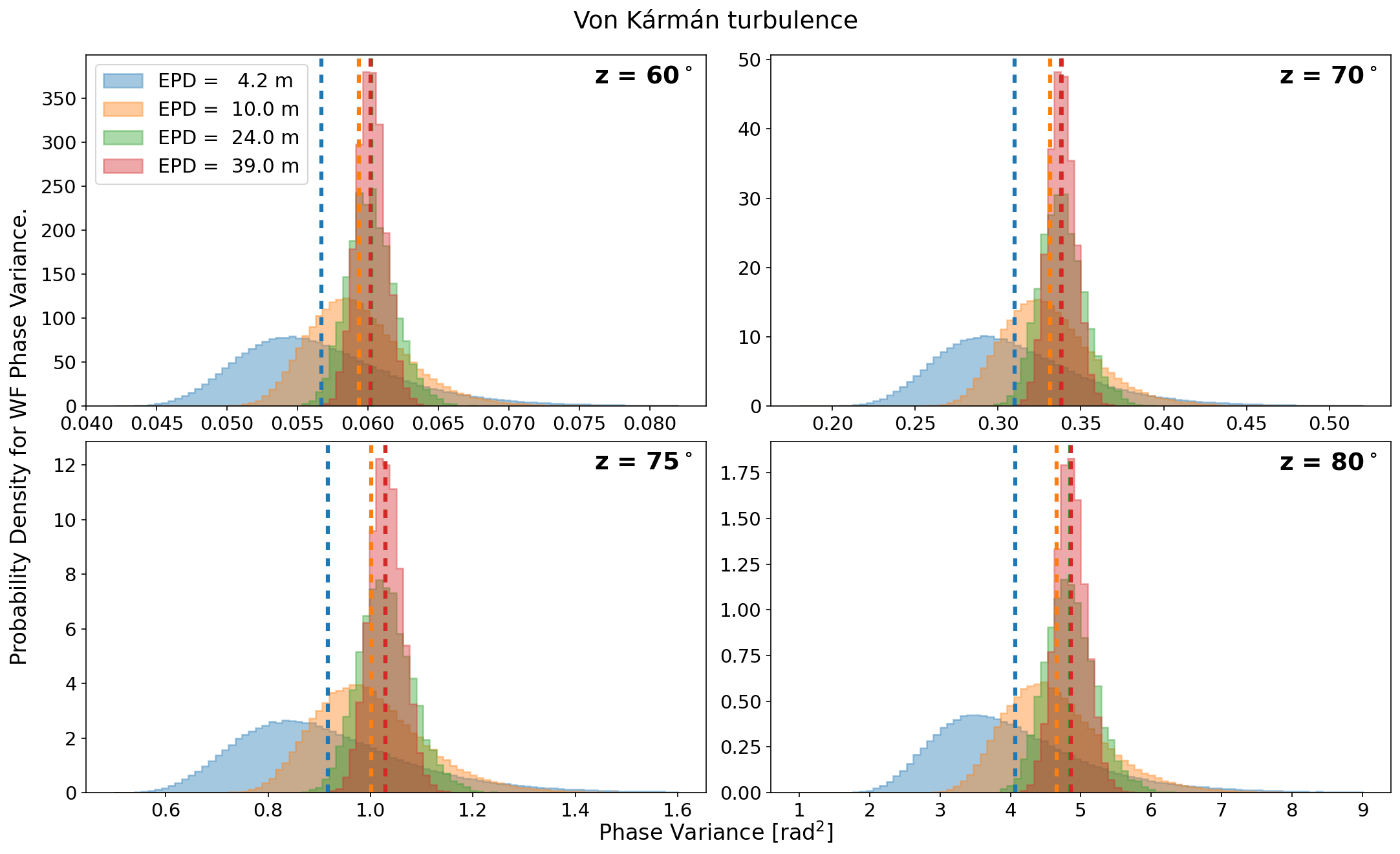}
    \end{subfigure}
    \hfill
    \begin{subfigure}{0.495\textwidth}     % Use {0.495\textwidth} for non-REFEREE version
        \centering
        \includegraphics[width=\textwidth]{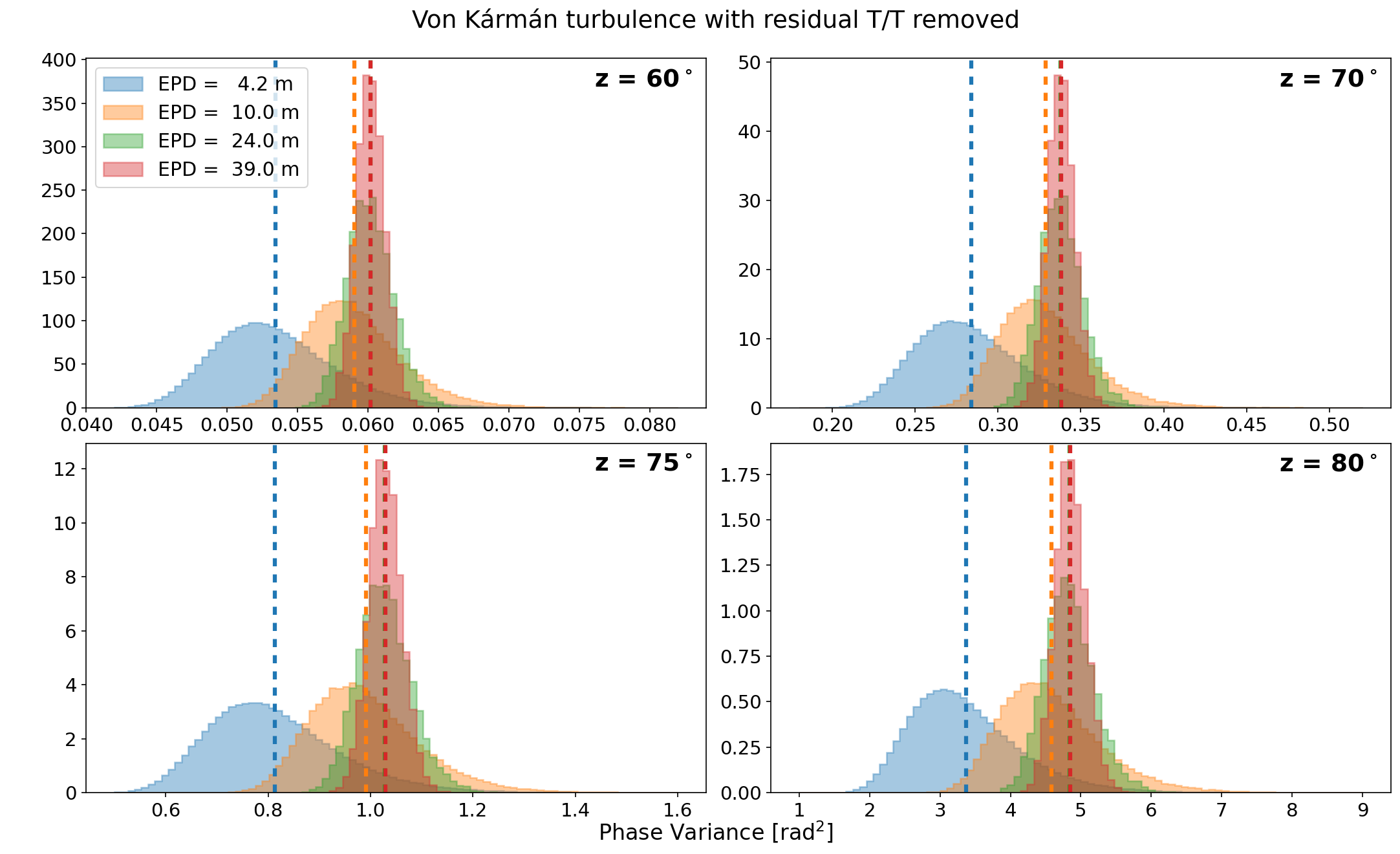}
    \end{subfigure}
\caption{Probability density function (PDF) estimators represented by area-normalized histograms of the CA-induced wavefront phase variance at $\lambda_{\rm Sci} = 1.25\,\mu\text{m}$ ($\lambda_{\rm WFS} = 589.16$~nm). The columns isolate four distinct zenith distance scenarios ($\zeta = 60^\circ, 70^\circ, 75^\circ,$ and $80^\circ$). The top row maps Kolmogorov turbulence, while the bottom row maps von K\'arm\'an turbulence ($L_0 = 25$~m), with the left and right sides showing the data before and after filtering out the residual tip-tilt modes, respectively. Color-coded distributions distinguish the telescope aperture scales (EPD = 4.2, 10, 24, and 39~m). Vertical dashed lines mark the statistical mean values computed directly from the Monte Carlo simulation pool ($\simeq 6400$ realizations for larger scales, extending up to $\sim 5\times10^{5}$ for the 4.2~m aperture). Note the transition from highly skewed, asymmetric profiles at smaller apertures to narrower, symmetric Gaussian-like behaviors as the EPD approaches or exceeds the outer scale $L_0$.}
    \label{fig:histo_sg2}
\end{sidewaysfigure*}

\begin{sidewaysfigure*} 
    \centering
    \begin{subfigure}{0.495\textwidth}     % Use {0.495\textwidth} for non-REFEREE version
        \centering
        \includegraphics[width=\textwidth]{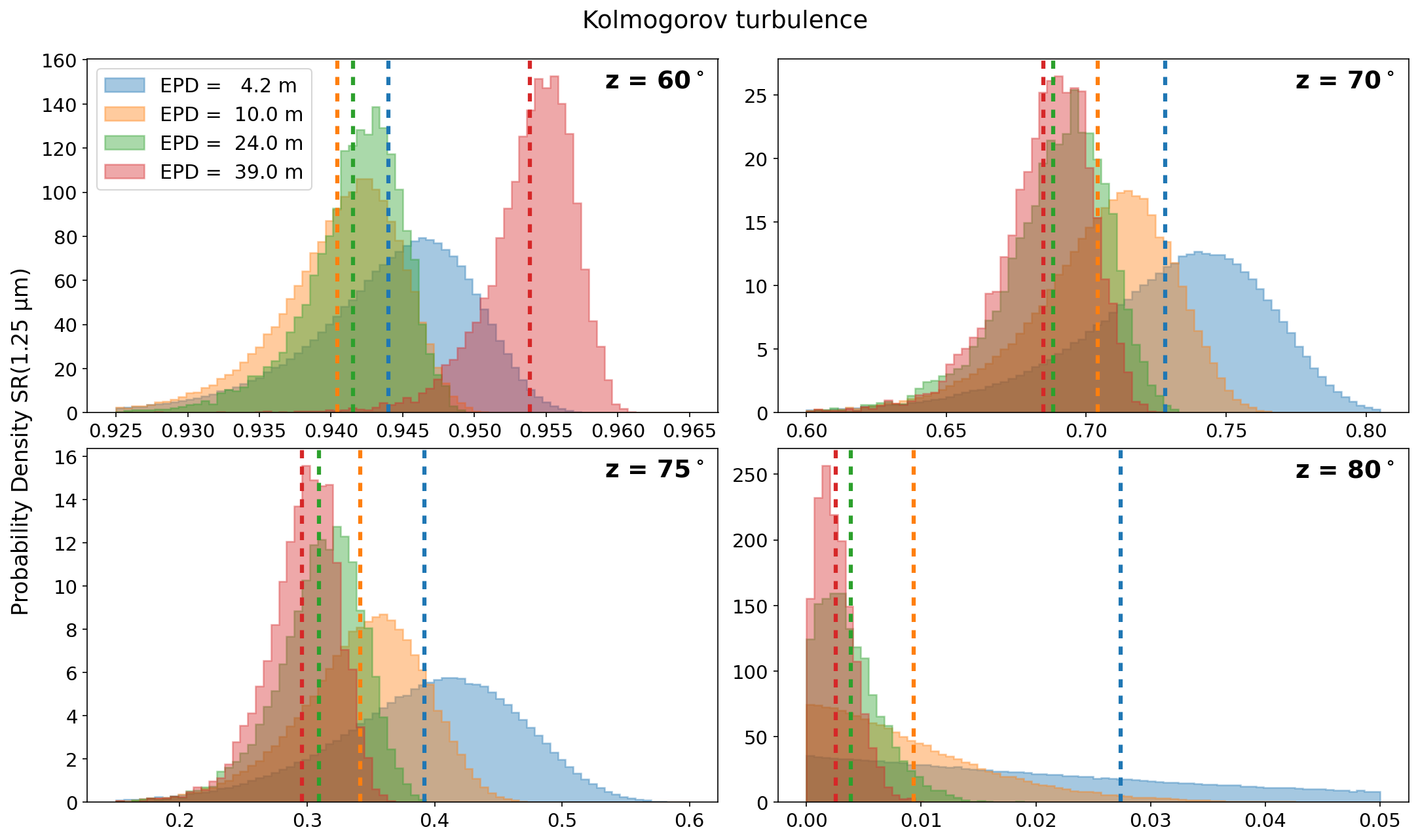}
    \end{subfigure}
    \hfill
    \begin{subfigure}{0.495\textwidth}     % Use {0.495\textwidth} for non-REFEREE version
        \centering
        \includegraphics[width=\textwidth]{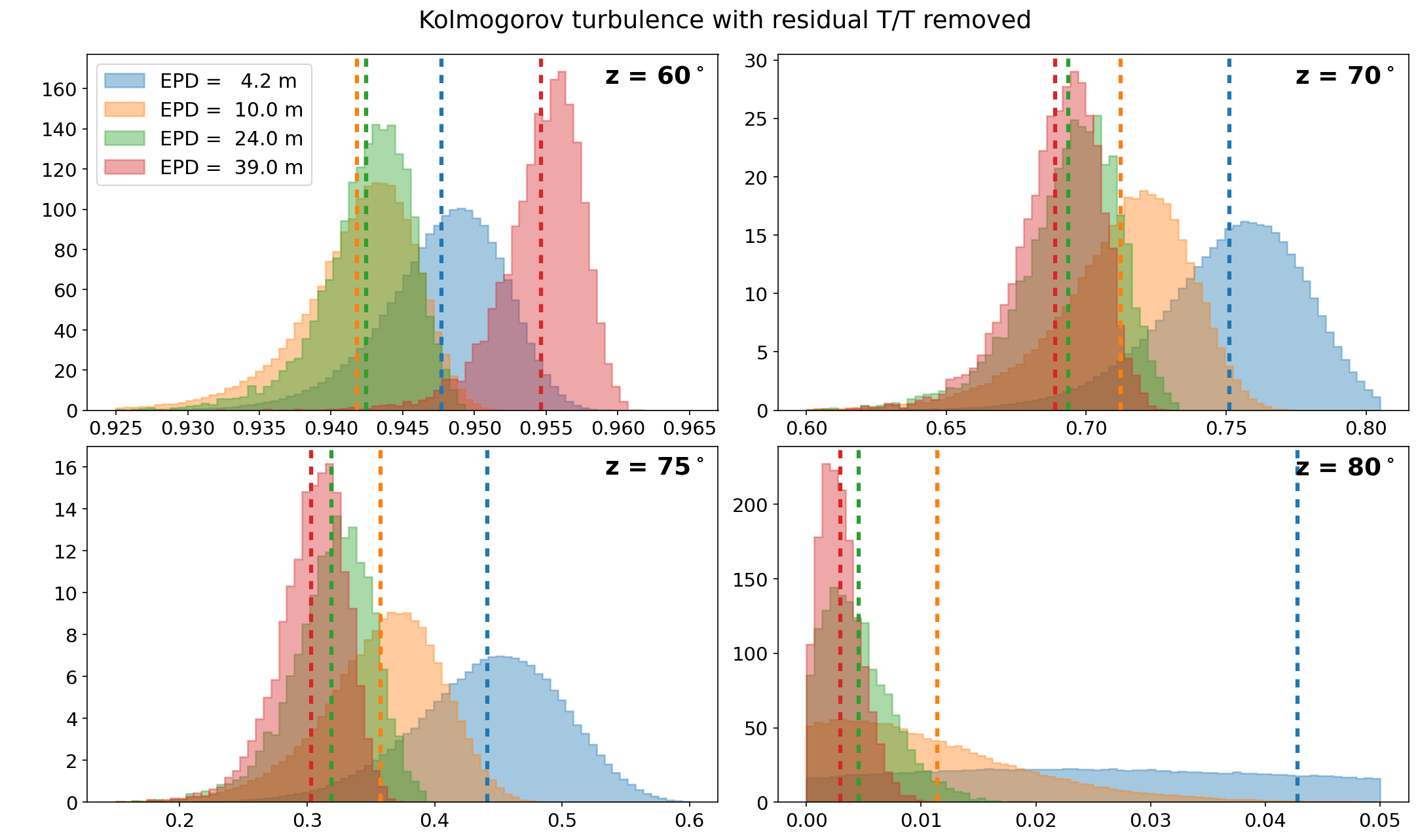}
    \end{subfigure}

    \begin{subfigure}{0.495\textwidth}     % Use {0.495\textwidth} for non-REFEREE version
        \centering
        \includegraphics[width=\textwidth]{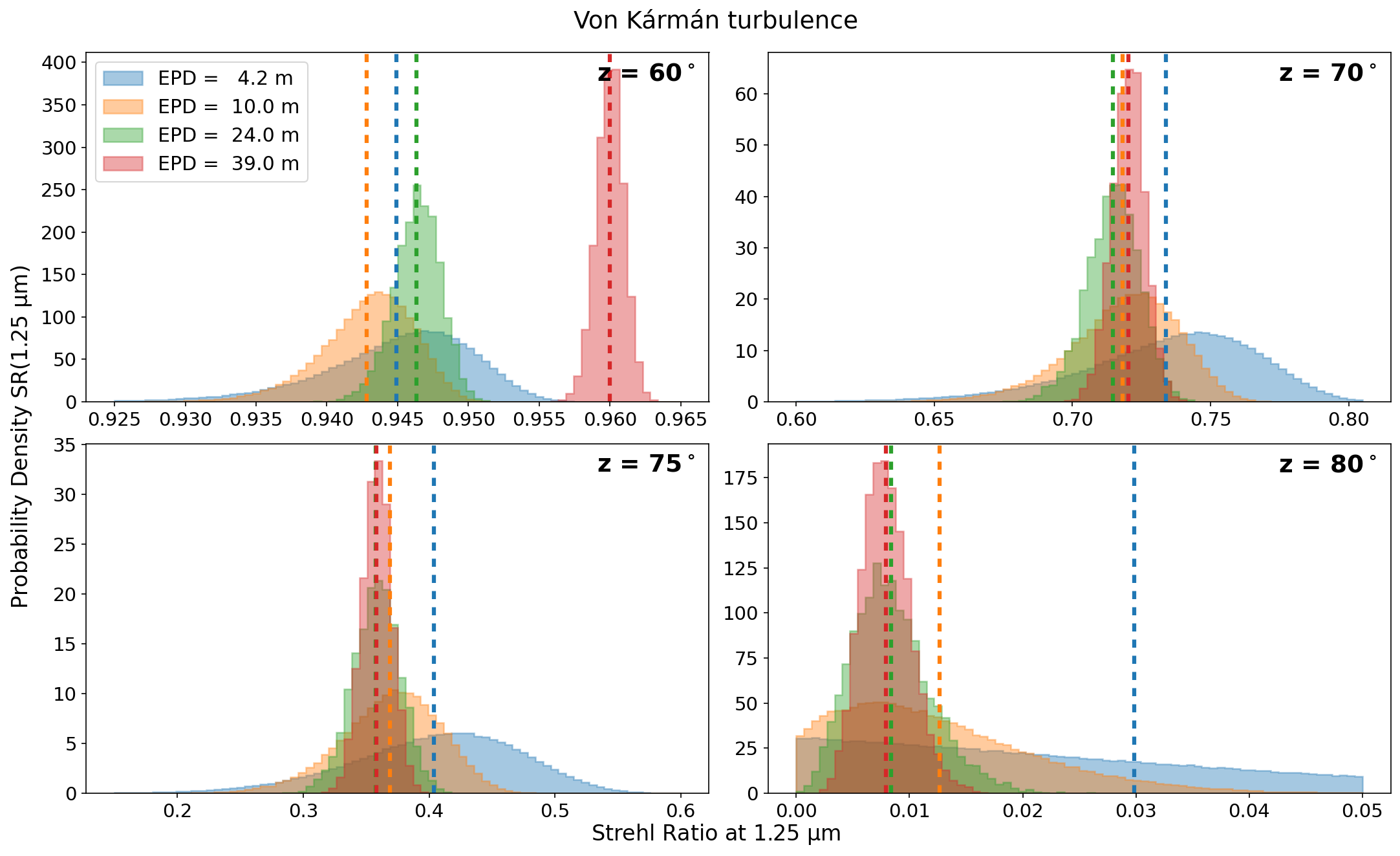}
    \end{subfigure}
    \hfill
    \begin{subfigure}{0.495\textwidth}     % Use {0.495\textwidth} for non-REFEREE version
        \centering
        \includegraphics[width=\textwidth]{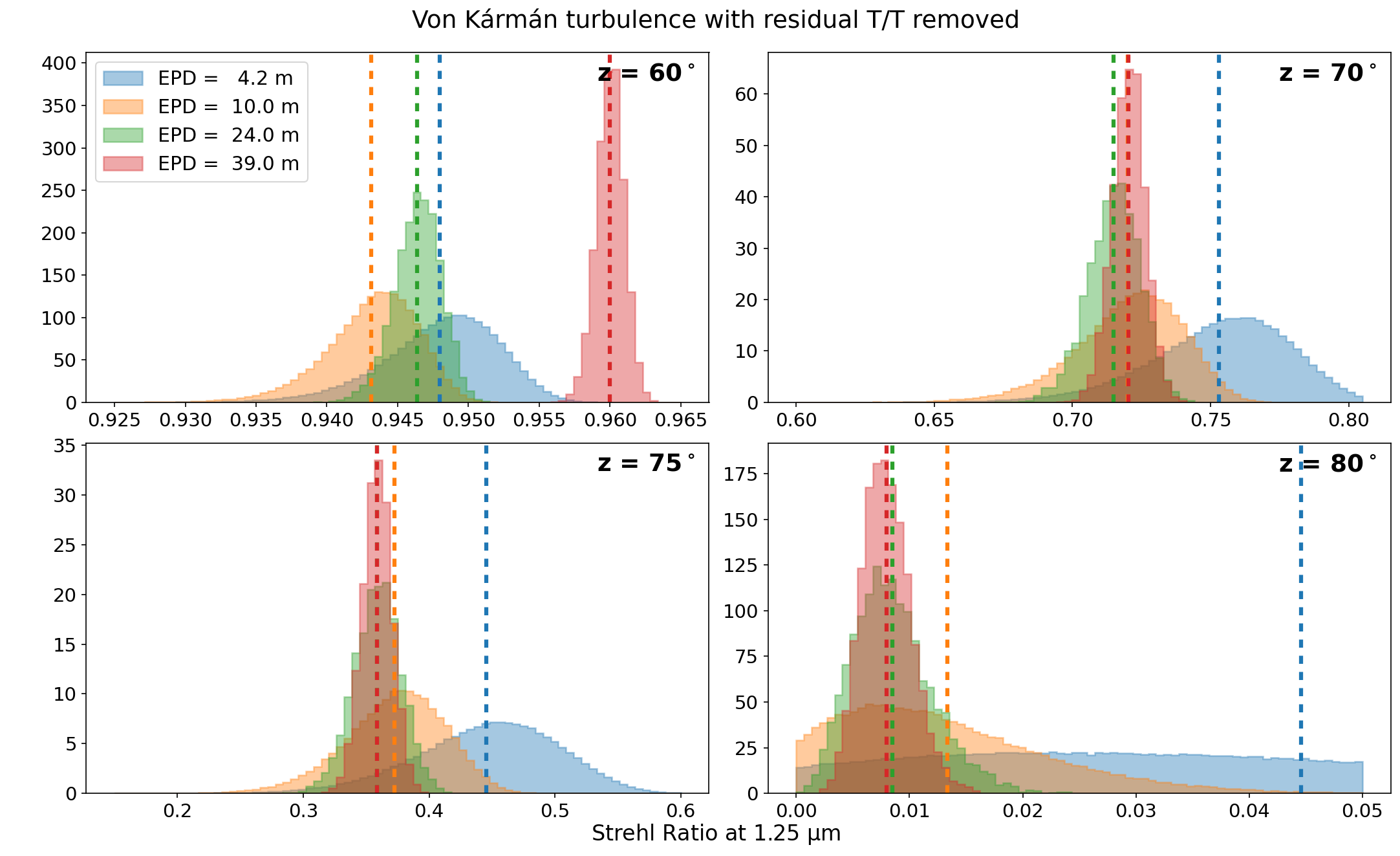}
    \end{subfigure}
\caption{Probability density function (PDF) estimators represented by area-normalized histograms of the Strehl ratio (SR) evaluated at $\lambda_{\rm Sci} = 1.25\,\mu\text{m}$ ($\lambda_{\rm WFS} = 589.16$~nm) under the influence of chromatic anisoplanatism. The layout structure mirrors Fig. 7, isolating distinct zenith distances ($\zeta = 60^\circ, 70^\circ, 75^\circ,$ and $80^\circ$) across columns. The top row maps Kolmogorov conditions and the bottom row presents von K\'arm\'an turbulence ($L_0 = 25$~m), with the left and right columns displaying the data before and after filtering the CA-induced tip-tilt modes, respectively. Color-coded profiles mark the individual aperture scales (EPD = 4.2, 10, 24, and 39~m). Vertical dashed lines indicate the statistical mean values computed from the Monte Carlo simulation pool. The distributions exhibit marked negative skewness and high kurtosis, displaying prolonged tails toward low-SR, high-error events that become exceptionally broad for the smallest telescope aperture (EPD = 4.2~m) at extreme zenith angles.}
    \label{fig:histo_SR}
\end{sidewaysfigure*} 

In Fig.~\ref{fig:boxplot_sg2_Kolmo} and Fig.~\ref{fig:boxplot_sg2_VK} we show, for the Kolmogorov and \VK~cases respectively,  the phase variance obtained from the numerical simulations for $\lW=0.58916~\mu$m and $\lS=1.25~\mu$m and four telescope apertures (EPD=4.2, 10, 24 and 39~m).  The phase variance as a function of zenith angle ($\zeta$) closely follows the theoretical expressions in Eq.~\eqref{eq:sg2_OPD_CA_1}. This is true for all EPDs considered, although the theoretical expression slightly overestimates the phase variance for the smallest aperture  at the highest zenith angles considered. One possible explanation may be how the PSs are numerically generated (e.g. missing sub-pixel spatial frequencies as discussed in Section~\ref{subsec:PS_generation}). Filtering out the tip-tilt components of the residual wavefront leads to only a slight reduction in phase variance at $\lS$. 

The theoretical CA expression for phase variance gives an overestimate in the case of \VK~turbulence, and the difference increases with telescope aperture. This is not an artefact of the PS generation but a true effect due to how a finite outer scale reduces wavefront variance; this reduction effect increases as the telescope aperture approaches and even exceeds the outer scale. However, when we consider a more benign scenario with $r_0^{\text{Kolmo}}=0.183$~m the theoretical Kolmogorov values slightly underestimate the  \VK~simulation results. This indicates that when computing the mean phase variance the role of $L_0=25$~m is minor and the major contribution is the $r_0$ to use for a fair comparison (see our comment in Section~\ref{subsec:PS_generation} about using $r_0^{\text{Kolmo}}(0.5\, \mu m)=0.183$~m). 

\begin{table}[ht!]
\caption{Percentage reduction in the CA phase variance when removing the tip-tilt component for the Kolmogorov (boldface) and \VK~cases.}
\label{table:tt_reduction}
\centering 
\begin{tabular}{l c c c c } 
\hline\hline
 $\zeta$      &  D= 4.2 m            & D= 10 m            &   D= 24 m            & D= 39 m             \\  \hline
$60^{\circ}$  &  \textbf{-6.0}/-5.2  & \textbf{-2.3}/-0.5 &  \textbf{-1.4}/-0.1  & \textbf{-1.2}/-0.03  \\       
$70^{\circ}$  &  \textbf{-8.8}/-7.7  & \textbf{-3.2}/-0.8 &  \textbf{-1.9}/-0.1  & \textbf{-1.6}/-0.05  \\       
$75^{\circ}$  &  \textbf{-12.}/-10   & \textbf{-4.2}/-1.1 &  \textbf{-2.5}/-0.2  & \textbf{-2.0}/-0.06  \\       
$80^{\circ}$  &  \textbf{-18.}/-16   & \textbf{-6.0}/-1.5 &  \textbf{-3.4}/-0.3  & \textbf{-2.8}/-0.08  \\ \hline
\end{tabular}
\end{table}

In both the Kolmogorov and \VK~scenarios we notice how small the effect of removing the tip-tilt components of the CA residual phase are. Table ~\ref{table:tt_reduction} shows the percentage reduction of phase variance as a function of zenith angle for the different pupil diameters considered and the Kolmogorov (in boldface) and \VK~cases. The reduction in CA phase variance due to removing the tip-tilt component  decreases with increase in telescope aperture, as would be expected, and increases with zenith angle. The relevance of the CA-induced tip-tilt component becomes extremely marginal (around 1\% or smaller) for apertures above 10 m in the \VK~case. This is consistent with our interpretation that the CA error is dominated by high spatial frequencies and  the presence of a finite $L_0$ causes saturation effects as the telescope aperture approaches $L_0$. Visually we see this high frequency content in the example of the residual wavefront shown in Fig.~\ref{fig:wf_example}.

The Strehl Ratio (SR) as a function of zenith distance, $\zeta$, and aperture diameter is shown in Fig.~\ref{fig:boxplot_SR}. The trends agree with what observed with the phase variance values in Fig.~\ref{fig:boxplot_sg2_Kolmo} and Fig.~\ref{fig:boxplot_sg2_VK} but the SR values provide a better understanding of the effect on the final image quality. The effect of removing tip-tilt is also very small on the SR values. 

Although the reductions in SR due to CA are moderate at the lowest zenith distances considered (60$^\circ$), they may be important for extreme AO systems such as those intended for the detection of exoplanets. At large zenith angles the reduction in SR is severe and the effect increases with telescope aperture, especially in Kolmogorov turbulence. While astronomical observations at $\zeta=80^\circ$ are very unusual we present here this scenario to illustrate how quickly the error increases with $\zeta$ and as a warning in the field of FSOC where it is not unusual to acquire and track LEOs at elevations as low as $5^\circ$.

The histograms in Fig.~\ref{fig:histo_sg2} show the distribution of  CA phase variance values for the four EPDs and a subset of the $\zeta$ values explored in Fig.~\ref{fig:boxplot_sg2_Kolmo}, Fig.~\ref{fig:boxplot_sg2_VK} and Fig.~\ref{fig:boxplot_SR}. The histograms are normalised to unity area and given the large sample of values they should be considered as good representation of the underlying Probability Density Functions (PDF) for the phase variance due to CA. The histograms for the Kolmogorov and \VK~scenarios, with and without tip-tilt filtered out, exhibit large positive skewness and high kurtosis; that is, they have long tails towards high residual variance values with rare events happening more often than would be expected in the case of normal distributions. This is especially true for the smallest telescope aperture at EPD=4.2~m  which exhibits broader and more asymmetrical phase variance distributions than large telescope apertures. For all cases of (EPD, $\zeta$) combinations the phase variance histograms with or without tip-tilt filtered out are nearly indistinguishable from each other, again reflecting how little content of CA-induced error is associated with the tip-tilt modes.  

From these observations we notice there will be a large number of instances where the phase variance is much higher than the mean or the value determined from the theoretical expression. The histograms basically shift towards higher phase variance values as the zenith angle increases, and the effect of removing tip-tilt is very small. The shape of the histograms are very different in the case of \VK~turbulence; overall the histograms for all EPDs appear to be spread over significantly narrower ranges but for the cases of EPD=24~m and EPD=39~m their phase variance distributions become much narrower and more symmetrical, looking like Gaussian distributions. We associate this behavior to a saturation effect when the EPD approaches or exceeds the value for $L_0$ in \VK~turbulence.
 
Fig.~\ref{fig:histo_SR} shows the histograms of SR reduction due to CA for the four EPDs and the same subset of $\zeta$ values as in Fig.~\ref{fig:histo_sg2}. Again we notice  asymmetrical SR distributions but with negative skewness (as expected e.g. from the Mar\'echal approximation relating phase variance to SR, valid for SR $\gtrsim 0.15\%$)  and high kurtosis. Again both in the Kolmogorov and  \VK~cases the larger the EPD the narrower and more symmetrical the SR distributions; an extreme case results with EPD=4.2~m case at $\zeta=80^\circ$ with an SR distribution which decays very slowly over a broad range of values. As with the phase variance histograms,  we only notice a  minor change with and without tip-tilt filtering applied indicative of the little relavance of the CA-induced tip-tilt modes. 

An interesting observation arises when looking in detail at the results for the SR distributions at $\zeta = 60^\circ$. It may be surprising when comparing the distributions of phase variance for the different telescope apertures for the same $\zeta = 60^\circ$ in Fig.~\ref{fig:histo_sg2}: while the phase variance is highest for EPD=39~m,  yet the SR is also highest for this aperture. We interpret this as the phase variance values being very small with some spatial correlation, contrary to the assumptions of the Mar\'{e}chal approximation. This was tested with numerical simulations where white\footnote{Assumption in the Mar\'{e}chal approximation.} Gaussian phase distribution of a very small variance  was compared with the same small variance of non-white Gaussian phase distribution (i.e. exhibiting some small degree of spatial correlation) - the SR computed with the Mar\'{e}chal approximation slightly underestimates the SR value computed from  its mathematical definition, Eq.~\eqref{eq:SR_def}.

%%%%%%%%%%%%%%%%%%%%%%%%%%%%%%%%%%%%%%%%%%%%%%%%%%%%%%%%%%%%%%
\section{Conclusions}
\label{sec:conclusion}

In this paper, we have carried out an exhaustive comparison between the theoretical expressions for wavefront phase variance caused by Chromatic Anisoplanatism (CA) and a Monte Carlo simulation framework independent of any approximations  in the theory assuming the near-field approximation to describe the optical propagation through atmospheric turbulence. To our knowledge, such a detailed comparison has not yet been carried out.
 
The wavefront phase variance values computed through the simulations are in excellent agreement with the theoretical expressions. Beyond supporting the validity of the theory, our numerical simulations provide a powerful mechanism to obtain additional information by allowing to access first and second-order statistics for the phase variance and the Strehl Ratio (SR).

Our results indicate the severity of CA for large zenith distances $\zeta \gtrsim 70^\circ$,  for which the SR loss exceeds 30\% for all aperture diameters and Kolmogorov and \VK~scenarios. We have also found that the residual wavefront due to CA is largely dominated by high spatial frequencies. In consequence, CA-induced tip-tilt modes represent a minor fraction of the total CA wavefront phase variance and an independent tip-tilt compensation would not result in a  substantial benefit, e.g. in an LGS-based scenario the use of an NGS for the overall tip-tilt determination would not mitigate the CA effects.

In the case of Kolmogorov turbulence the wavefront phase variance always increases with both the zenith distance ($\zeta$)  and telescope  diameter, e.g. at a given $\zeta$ a 39~m diameter always exhibits larger phase variance than a 24~m telescope. This trend does not happen in \VK~turbulence where at large $\zeta$ the phase variance saturates for telescope diameters larger than the outer scale $L_0$ and exhibits nearly the same phase variance for a 24~m and 39~m telescope diameters. 

In the Kolmogorov case the distributions of wavefront phase variance values are asymmetric and the larger the aperture the narrower the  distributions. In the \VK~case the distributions become narrower with increasing telescope diameter and  become symmetric when approaching or exceeding $L_0$. These behaviours do not change when the tip-tilt components due to CA are filtered out. The SR histograms are consistent with the trends observed with the distribution of phase variance values in terms of asymmetry and width. There is an exception occuring at $\zeta = 60^\circ$ when despite the phase variance being higher for EPD=39~m, the SR is also the highest for this aperture. This is understood as a limitation of the Mar\'echal approximation.  

Finally, the simulation techniques and optimizations proposed in this work provide a highly efficient framework for future error-budgeting analyses. By significantly expanding the number of useful turbulence realizations without increasing storage overhead, this approach enables the robust statistical modeling required to characterize confidence levels for next-generation adaptive optics instruments.

%%%%%%%%%%%%%%%%%%%%%%%%%%%%%%%%%%%%%%%%%%%%%%%%%%%%%%%%%%%%%%
\begin{acknowledgements}
The authors acknowledge that this research was conducted independently and did not receive any specific grant from funding agencies in the public, commercial, or not-for-profit sectors. During the preparation of this manuscript, the authors used generative AI tools (Gemini and Claude) solely to enhance text readability, grammar, and figure caption formatting. All scientific content, numerical simulation datasets, and final text interpretations were generated and verified exclusively by the authors, who retain full responsibility for the integrity of the work.
\end{acknowledgements}

%%%%%%%%%%%%%%%%%%%%%%%%%%%%%%%%%%%%%%%%%%%%%%%%%%%%%%%%%%%%%%
% \bibliographystyle{aa}
% \bibliography{report}

\begin{thebibliography}{21}
\expandafter\ifx\csname natexlab\endcsname\relax\def\natexlab#1{#1}\fi

\bibitem[{Carbillet \& Riccardi(2010)}]{carbillet_numerical_2010}
Carbillet, M. \& Riccardi, A. 2010, Applied Optics, 49, G47

\bibitem[{Carbillet {et~al.}(2005)Carbillet, Vérinaud, Femenía, Riccardi, \&
  Fini}]{CAOS_2005}
Carbillet, M., Vérinaud, C., Femenía, B., Riccardi, A., \& Fini, L. 2005,
  Monthly Notices of the Royal Astronomical Society, 356, 1263

\bibitem[{Devaney \& Femen{\'\i}a-Castell{\'a}(2024)}]{d24}
Devaney, N. \& Femen{\'\i}a-Castell{\'a}, B. 2024, in Adaptive Optics Systems
  IX, Vol. 13097, SPIE, 158--169

\bibitem[{Devaney {et~al.}(2008)Devaney, Goncharov, \& Dainty}]{D08}
Devaney, N., Goncharov, A.~V., \& Dainty, J.~C. 2008, Appl. Opt., 47, 1072

\bibitem[{Ellerbroek {et~al.}(1994)Ellerbroek, Pompea, Robertson, \&
  Mountain}]{ellerbroek1994}
Ellerbroek, B.~L., Pompea, S.~M., Robertson, D.~J., \& Mountain, C.~M. 1994, in
  Proceedings of SPIE, Vol. 2201, Adaptive Optics in Astronomy, ed. M.~A. Ealey
  \& F.~Merkle (SPIE), 421--436

\bibitem[{Femen{\'\i}a-Castell{\'a} {et~al.}(2022)Femen{\'\i}a-Castell{\'a},
  Devaney, Cagigal, \& Bonaque-Gonz{\'a}lez}]{femenia2022}
Femen{\'\i}a-Castell{\'a}, B., Devaney, N., Cagigal, M.~N., \&
  Bonaque-Gonz{\'a}lez, S. 2022, in Adaptive Optics Systems VIII, Vol. 12185,
  SPIE, 1759--1773

\bibitem[{{Femen\'{\i}a Castell\'a} \& et~al.(2022)}]{FC22}
{Femen\'{\i}a Castell\'a}, B. \& et~al. 2022, in Adaptive Optics Systems VIII,
  ed. L.~Schreiber, D.~Schmidt, \& E.~Vernet, Vol. 12185, These proceedings

\bibitem[{Hyde {et~al.}(2026)Hyde, Spencer, \&
  Kalensky}]{hyde2026anisoplanatic}
Hyde, M.~W., Spencer, M.~F., \& Kalensky, M. 2026, Radio Science, 61,
  e2025RS008489

\bibitem[{Johnson {et~al.}(2020)Johnson, Johansson, Marino, Richards, Rimmele,
  Wang, \& Woeger}]{DKIST_20}
Johnson, L.~C., Johansson, E., Marino, J., {et~al.} 2020, in Adaptive {Optics}
  {Systems} {VII}, ed. D.~Schmidt, L.~Schreiber, \& E.~Vernet, Vol. 11448
  (Proc. SPIE), 114480T

\bibitem[{Kasten \& Young(1989)}]{Kasten1989}
Kasten, F. \& Young, A.~T. 1989, Applied Optics, 28, 4735

\bibitem[{Kolb {et~al.}(2015)Kolb, Marchetti, Schneller, Sarazin, \&
  Cirasuolo}]{ESO1}
Kolb, J., Marchetti, E., Schneller, D., Sarazin, M., \& Cirasuolo, M. 2015,
  Relevant Atmospheric Parameters for E-ELT AO Analysis and Simulations,
  Specification (SPE) ESO-258292, European Southern Observatory (ESO)

\bibitem[{Labriji {et~al.}(2022)Labriji, Herscovici-Schiller, \&
  Cassaing}]{L22}
Labriji, H., Herscovici-Schiller, O., \& Cassaing, F. 2022, Astronomy \&
  Astrophysics, 662, A61

\bibitem[{Motte {et~al.}(2024)Motte, Neichel, Fusco, F{\'e}tick, Sauvage,
  H{\'e}ritier, Ciss{\'e}, Blind, \& Lovis}]{Motte2024}
Motte, M., Neichel, B., Fusco, T., {et~al.} 2024, in Proc.\ SPIE, Vol. 13097,
  Adaptive Optics Systems {IX}, 130976A

\bibitem[{Nakajima(2006)}]{N06}
Nakajima, T. 2006, ApJ, 652, 1782

\bibitem[{Owner-Petersen \& Goncharov(2004)}]{O04}
Owner-Petersen, M. \& Goncharov, A.~V. 2004, in Ground-based Telescopes, ed.
  J.~M. Oschmann, Vol. 5489, 507--517

\bibitem[{{Quintero Noda} {et~al.}(2022){Quintero Noda}, {Schlichenmaier},
  {Bellot Rubio}, {L{\"o}fdahl}, {Khomenko}, {Jur{\v{c}}{\'a}k}, {Leenaarts},
  {Kuckein}, {Gonz{\'a}lez Manrique}, {Gun{\'a}r}, {Nelson}, {de la Cruz
  Rodr{\'\i}guez}, {Tziotziou}, {Tsiropoula}, {Aulanier}, {Aboudarham},
  {Allegri}, {Alsina Ballester}, {Amans}, {Asensio Ramos}, {Bail{\'e}n},
  {Balaguer}, {Baldini}, {Balthasar}, {Barata}, {Barczynski}, {Barreto
  Cabrera}, {Baur}, {B{\'e}chet}, {Beck}, {Bel{\'\i}o-As{\'\i}n},
  {Bello-Gonz{\'a}lez}, {Belluzzi}, {Bentley}, {Berdyugina}, {Berghmans},
  {Berlicki}, {Berrilli}, {Berkefeld}, {Bettonvil}, {Bianda}, {Bienes
  P{\'e}rez}, {Bonaque-Gonz{\'a}lez}, {Braj{\v{s}}a}, {Bommier}, {Bourdin},
  {Burgos Mart{\'\i}n}, {Calchetti}, {Calcines}, {Calvo Tovar}, {Campbell},
  {Carballo-Mart{\'\i}n}, {Carbone}, {Carlin}, {Carlsson}, {Castro L{\'o}pez},
  {Cavaller}, {Cavallini}, {Cauzzi}, {Cecconi}, {Chulani}, {Cirami},
  {Consolini}, {Coretti}, {Cosentino}, {C{\'o}zar-Castellano}, {Dalmasse},
  {Danilovic}, {De Juan Ovelar}, {Del Moro}, {del Pino Alem{\'a}n}, {del Toro
  Iniesta}, {Denker}, {Dhara}, {Di Marcantonio}, {D{\'\i}az Baso}, {Diercke},
  {Dineva}, {D{\'\i}az-Garc{\'\i}a}, {Doerr}, {Doyle}, {Erdelyi}, {Ermolli},
  {Escobar Rodr{\'\i}guez}, {Esteban Pozuelo}, {Faurobert}, {Felipe}, {Feller},
  {Feijoo Amoedo}, {Femen{\'\i}a Castell{\'a}}, {Fernandes}, {Ferro
  Rodr{\'\i}guez}, {Figueroa}, {Fletcher}, {Franco Ordovas}, {Gafeira},
  {Gardenghi}, {Gelly}, {Giorgi}, {Gisler}, {Giovannelli}, {Gonz{\'a}lez},
  {Gonz{\'a}lez}, {Gonz{\'a}lez-Cava}, {Gonz{\'a}lez Garc{\'\i}a},
  {G{\"o}m{\"o}ry}, {Gracia}, {Grauf}, {Greco}, {Grivel}, {Guerreiro},
  {Guglielmino}, {Hammerschlag}, {Hanslmeier}, {Hansteen}, {Heinzel},
  {Hern{\'a}ndez-Delgado}, {Hern{\'a}ndez Su{\'a}rez}, {Hidalgo}, {Hill},
  {Hizberger}, {Hofmeister}, {J{\"a}gers}, {Janett}, {Jarolim}, {Jess},
  {Jim{\'e}nez Mej{\'\i}as}, {Jolissaint}, {Kamlah}, {Kapit{\'a}n},
  {Ka{\v{s}}parov{\'a}}, {Keller}, {Kentischer}, {Kiselman}, {Kleint},
  {Klvana}, {Kontogiannis}, {Krishnappa}, {Ku{\v{c}}era}, {Labrosse}, {Lagg},
  {Landi Degl'Innocenti}, {Langlois}, {Lafon}, {Laforgue}, {Le Men}, {Lepori},
  {Lepreti}, {Lindberg}, {Lilje}, {L{\'o}pez Ariste}, {L{\'o}pez
  Fern{\'a}ndez}, {L{\'o}pez Jim{\'e}nez}, {L{\'o}pez L{\'o}pez}, {Manso
  Sainz}, {Marassi}, {Marco de la Rosa}, {Marino}, {Marrero}, {Mart{\'\i}n},
  {Mart{\'\i}n G{\'a}lvez}, {Mart{\'\i}n Hernando}, {Masciadri}, {Mart{\'\i}nez
  Gonz{\'a}lez}, {Matta-G{\'o}mez}, {Mato}, {Mathioudakis}, {Matthews}, {Mein},
  {Merlos Garc{\'\i}a}, {Moity}, {Montilla}, {Molinaro}, {Molodij}, {Montoya},
  {Munari}, {Murabito}, {N{\'u}{\~n}ez Cagigal}, {Oliviero}, {Orozco
  Su{\'a}rez}, {Ortiz}, {Padilla-Hern{\'a}ndez}, {Pa{\'e}z Ma{\~n}{\'a}},
  {Paletou}, {Pancorbo}, {Pastor Ca{\~n}edo}, {Pastor Yabar}, {Peat},
  {Pedichini}, {Peixinho}, {Pe{\~n}ate}, {P{\'e}rez de Taoro}, {Peter},
  {Petrovay}, {Piazzesi}, {Pietropaolo}, {Pleier}, {Poedts}, {P{\"o}tzi},
  {Podladchikova}, {Prieto}, {Quintero Nehrkorn}, {Ramelli}, {Ramos Sapena},
  {Rasilla}, {Reardon}, {Rebolo}, {Regalado Olivares}, {Reyes
  Garc{\'\i}a-Talavera}, {Riethm{\"u}ller}, {Rimmele}, {Rodr{\'\i}guez
  Delgado}, {Rodr{\'\i}guez Gonz{\'a}lez}, {Rodr{\'\i}guez-Losada},
  {Rodr{\'\i}guez Ramos}, {Romano}, {Roth}, {Rouppe van der Voort}, {Rudawy},
  {Ruiz de Galarreta}, {Ryb{\'a}k}, {Salvade}, {S{\'a}nchez-Capuchino},
  {S{\'a}nchez Rodr{\'\i}guez}, {Sangiorgi}, {Say{\`e}de}, {Scharmer},
  {Scheiffelen}, {Schmidt}, {Schmieder}, {Scir{\`e}}, {Scuderi}, {Siegel},
  {Sigwarth}, {Sim{\~o}es}, {Snik}, {Sliepen}, {Sobotka}, {Socas-Navarro},
  {Sola La Serna}, {Solanki}, {Soler Trujillo}, {Soltau}, {Sordini}, {Sosa
  M{\'e}ndez}, {Stangalini}, {Steiner}, {Stenflo}, {{\v{S}}t{\v{e}}p{\'a}n},
  {Strassmeier}, {Sudar}, {Suematsu}, {S{\"u}tterlin}, {Tallon}, {Temmer},
  {Tenegi}, {Tritschler}, {Trujillo Bueno}, {Turchi}, {Utz}, {van Harten}, {van
  Noort}, {van Werkhoven}, {Vansintjan}, {Vaz Cedillo}, {Vega Reyes}, {Verma},
  {Veronig}, {Viavattene}, {Vitas}, {V{\"o}gler}, {von der L{\"u}he},
  {Volkmer}, {Waldmann}, {Walton}, {Wisniewska}, {Zeman}, {Zeuner}, {Zhang},
  {Zuccarello}, \& {Collados}}]{QuinteroC:EST}
{Quintero Noda}, C., {Schlichenmaier}, R., {Bellot Rubio}, L.~R., {et~al.}
  2022, \aap, 666, A21

\bibitem[{Rimmele \& Marino(2011)}]{RM11}
Rimmele, T.~R. \& Marino, J. 2011, {Living} {Rev.} {Solar} {Physics}, 8

\bibitem[{Sasiela(1992)}]{S92}
Sasiela, R.~J. 1992, J. Opt. Soc. Am., 9, 1398

\bibitem[{Wallner(1976)}]{W76}
Wallner, E.~P. 1976, in Proc. SPIE, Vol.~75, Imaging Through the Atmosphere,
  ed. J.~C. Wyant, 119--125

\bibitem[{Wallner(1977)}]{W77}
Wallner, E.~P. 1977, J. Opt. Soc. Am., 67, 407

\bibitem[{Wallner(1984)}]{W84}
Wallner, E.~P. 1984, J. Opt. Soc. Am. A, 1, 785

\end{thebibliography}

%%%%%%%%%%%%%%%%%%%%%%%%%%%%%%%%%%%%%%%%%%%%%%%%%%%%%%%%%%%%%%%
\begin{appendix}
\nolinenumbers
\label{App:Theory}
\onecolumn

\section{Theoretical expressions for the OPD variance due to Chromatic Anisoplanatism}

Following the formalism in \citet{W84} and upon correcting some typos, the expression for the variance of the Optical Path Difference (OPD) at  wavelength $\lA$ when correction is derived from  wavelength $\lB$ is given by the following set of equations: 

\begin{align}
\sigma^2_{OPD}\left(\lA,\lB \right) & =  2.914 \cdot \am \int_0^H  \mathrm{d}h \cdot \Cn \left\{ \left|\vec{s}(\vec{0}, \lA, \lB, h)\right|^{5/3} + \iint\limits_{-\infty}^{\infty} \mathrm{d}\vec{\rho} \cdot \mathscr{A}(\vec{\rho}) \left[ \left|\vec{s}(\vec{\rho}, \lA, \lA, h)\right|^{5/3} - \left|\vec{s}(\vec{\rho}, \lA, \lB, h)\right|^{5/3} \right] \right\} \label{eq:sg2_OPD_CA_1} \\
      \vec{s}(\vec{y}, \lA, \lB, h) & = \left( 1-\frac{h}{H}\right)\cdot \vec{y} + \Delta b_0(\lA, \lB) \cdot \left[1-\frac{P(h)}{P_0} - \left(1- \frac{P(H)}{P_0} \right)\frac{h}{H}\right] \cdot \hat{u}_b  \label{eq:sg2_OPD_CA_2}\\
       W(\vec{x}) & =
    \begin{cases}
      1 / \left(\pi R^2\right) & \text{if } |\vec{x}| \leq R\\
      0, & \text{otherwise}
    \end{cases}   \label{eq:sg2_OPD_CA_3}  \\   
     \mathscr{A}(\vec{\rho}) & = \iint\limits_{-\infty}^{\infty} \mathrm{d}\vec{x} \cdot W\left(\vec{x} + \frac{\vec{\rho}}{2}\right) \cdot W\left(\vec{x} -\frac{\vec{\rho}}{2}\right) = 
         \begin{cases}
           \frac{2}{(\pi R)^2} \left[\arccos\left(\frac{\rho}{2\cdot R}\right) -  \frac{\rho}{2\cdot R}\sqrt{1 - \left( \frac{\rho}{2\cdot                       R}\right)^2} \right]   &  \text{if } \rho \leq 2\cdot R\\
            0,  & \text{otherwise}
        \end{cases}   \label{eq:sg2_OPD_CA_4}     
\end{align}

\noindent where the different symbols and expressions are:  

\begin{description}[font=\normalfont\textendash\ \bfseries, 
                    leftmargin=\widthof{\textbullet\ $\vec{s}(\vec{y}, \lA, \lB, h)$:~}+0.5em, 
                    labelwidth=\widthof{\textbullet\ $\vec{s}(\vec{y}, \lA, \lB, h)$:~},                                                             
                    style=sameline]                                                        
 
  \item[$\vec{y},\, \vec{x}, \, \vec{\rho}$:] are 2D-vectors in the pupil and/or beam footprints on planes perpendicular to the line-of-sight. 

  \item[$\vec{s}(\vec{y}, \lA, \lB, h)$:]is a 2D vector with the lateral separation at the plane perpendicular to the line-of-sight at height $h$ between rays of wavelength $\lA$ and $\lB$ which enter the atmosphere at the same point and are separated by $\vec{y}$\ at the telescope aperture (e.g. $h=0$).
  
  \item[$\am$:] is the airmass at zenith distance $\zeta$. For zenith distances smaller than $65^\circ$ the usual approximation $\am \simeq \sec(\zeta)$ is adequate with no appreciable errors. 

  \item[$\Delta b_0(\lA, \lB)$:] is the lateral shift at the telescope aperture  of two rays at wavelengths $\lA$ and $\lB$ which enter the atmosphere at the same point. \citet{W76} finds the expression for $\Delta b_0$ assuming a plane-parallel astmophere but we consider the spherical geometry atmosphere derived in \citet{L22}.

  \item[$\hat{u}_b$:] is the unitary vector along the dispersion direction on the plane perpendicular to the line-of-sight at  altitude $h$.

  \item[$H$:] is the distance to the target (e.g. the LGS or the NGS or the Solar surface). 

  \item[$P_0$:] is the atmospheric pressure at the telescope site and $P(h)$ the  pressure at a height $h$ above the telescope site. 

  \item[$W(\vec{x})$:] is the area-normalized entrance pupil function. The second part in Eq.~\eqref{eq:sg2_OPD_CA_3} is for the case of an unobstructed circular pupil.

  \item[$\mathscr{A}(\vec{\rho})$:] is the overlapping function between two pupils separated by $\vec{\rho}$. The second part in Eq.~\eqref{eq:sg2_OPD_CA_4} is for the case of unobstructed circular pupils.
\end{description}

It is interesting to notice that due to Eq.\eqref{eq:sg2_OPD_CA_2} dependence on $\lA$ and $\lB$, $\vec{s}(\vec{y}, \lA, \lA, h)$ is independent of wavelength and takes on a very simple form: $\vec{s}(\vec{y}, \lA, \lA, h)= (1-h/H)\vec{y}= \rho^{5/3}$, allowing to decouple the integrals in $h$ and pupil position $\vec{\rho}$ for that term. Then, for circular apertures and for a distant source (so that for all practical purposes $H=\infty$) it is possible to obtain a closed analytical expression for the term with $\vec{s}(\vec{y}, \lA, \lA, h)$ in Eq.\eqref{eq:sg2_OPD_CA_1}:
\begin{align}
I_2\left(\lA\right)  & =  2.914 \am \int_0^H  \mathrm{d}h \cdot \Cn \iint\limits_{-\infty}^{\infty} \mathrm{d}\vec{\rho} \cdot \left|\vec{s}(\vec{\rho}, \lambda_a, \lambda_a, h)\right|^{5/3} \cdot \mathscr{A}(\vec{\rho}) \\
                     & =  2.914 \left[\sec(\zeta) \int_0^H  \mathrm{d}h \cdot C_n^2(h) \right] \cdot \left[ 2\pi  \int^{2R}_0 \mathrm{d}\rho \cdot \rho^{8/3}\mathrm{A}(\rho)\right] \simeq  0.05234 \left(\frac{D}{r_0(\lA;\zeta)}\right)^{5/3} \lambda^2  \label{eq:sg2_OPD_I2}
\end{align}
 
\noindent where $\mathrm{A}(\rho)$ is the 1D version of $\mathscr{A}(\vec{\rho})$ (i.e. $\mathrm{A}(\rho) = \mathrm{A}(\left|\vec{\rho}\right|) = \mathscr{A}(\vec{\rho})$ and $r_0(\lA;\zeta)$ is the Fried parameter at $\lA$ towards zenith distance $\zeta$. This closed form was used throughout the numerical evaluation of Eq.~\eqref{eq:sg2_OPD_CA_1} and the numerical simulations. 

As  discussed in Sec.~\ref{sec:theory}, the theoretical approach adopted by \citet{W84} makes three reasonable assumptions in the context of astronomical conditions where  moderate to weak atmospheric turbulence and large enough telescope apertures are found: 
\begin{enumerate}
\item Although Eq.\eqref{eq:sg2_OPD_CA_1} presents a continuous distribution of turbulence, \citet{W84} assumes no correlation between turbulence at two  distinct altitudes, e.g. turbulence layers at $h$ and $h+\text{d}h$ are totally independent. This matches perfectly with our numerical simulation where the atmosphere is distributed in discrete layers and the separation between those layers is on the order of hundreds or thousands of meters as depicted in Table~\ref{table:turb_profile} and Fig.~\ref{fig:turb_profile}. It is instructive for the following discussion to explicitely write the discrete version of Eq.~\eqref{eq:sg2_OPD_CA_1} for 8 independent turbulence layers becomes:
\begin{eqnarray}
\sigma^2_{OPD}\left(\lA,\lB \right) & = & I_1(\lA,\lB) + I_2(\lA) - I_3(\lA,\lB) \label{eq:sg2_discr_1}\\
I_1\left(\lA,\lB \right)  & = & 2.914 \cdot \am \cdot \sum_{i=1}^{8} C_{n,i}^2 \cdot \left|\vec{s}(\vec{0}, \lA, \lB, h_i)\right|^{5/3}\label{eq:sg2_discr_2} \\
I_2\left(\lA     \right)  & = & 2.914 \cdot \am \cdot \sum_{i=1}^{8} C_{n,i}^2 \cdot \iint\limits_{-\infty}^{\infty} \mathrm{d}\vec{\rho}\left|\vec{s}(\vec{\rho}, \lA, \lA, h_i)\right|^{5/3}\label{eq:sg2_discr_3} \\
I_3\left(\lA,\lB \right)  & = & 2.914 \cdot \am \cdot \sum_{i=1}^{8} C_{n,i}^2 \cdot \iint\limits_{-\infty}^{\infty} \mathrm{d}\vec{\rho}\left|\vec{s}(\vec{\rho}, \lA, \lB, h_i)\right|^{5/3}\label{eq:sg2_discr_4}
\end{eqnarray}

\item The theoretical derivation is based on the near-field approximation for the phase propagation of light through atmospheric turbulence. This approach is also referred to as geometrical propagation and consists of the addition of the phase along the propagation path thorugh the turbulence layers. By ignoring Fresnel diffraction between turbulence layers, the formalism is not able to account for amplitude fluctuations of the electromagnetic field (scintillation) nor additional changes in the phase as the light is collected at the telescope aperture. However such effects should not be considered a limitation of our approach because:\label{fresnel_justification} 
\begin{itemize}
\item AO astronomical observations are intended to be conducted mostly under moderate to weak turbulence conditions.
\item Scintillation effects are largely averaged out for the size of apertures considered in our simulations. 
\item Scintillation effects would only be relevant in the actual mechanism for the wavefront sensing but in this work we are implicitely assuming the wavefront sensing is perfect at $\lW$.
\end{itemize}

\item The theoretical analysis assumes Kolmogorov turbulence. In Sect.~\ref{sec:simulations} we  found strong evidence from the numerical simulations assuming \VK~turbulence that the assumption of a finite outer scale does not  impact the first-order statistical results (i.e. mean wavefront phase variance) with respect to the predictions from the theory. 

\item Not a limitation but a clarification: despite the fact that Eq.~\eqref{eq:sg2_OPD_CA_2} considers the height of the source above the observatory site, $H$, the formulation for the case of a finite distance (i.e. $H_{LGS}\sim 90$~Km) does not take into account the cone effect for LGS AO. 

\end{enumerate} 

\end{appendix}

\end{document}